\documentclass[acmsmall,screen,nonacm]{acmart}

\usepackage{booktabs} 
\usepackage{listings}
\usepackage{xcolor}
\usepackage{balance}
\usepackage{mathpartir}
\usepackage{amsmath, amssymb} 
\usepackage{enumitem} 

\lstdefinelanguage{yaml}{
  keywords={true,false,null,y,n},
  keywordstyle=\color{darkgray}\bfseries,
  sensitive=false,
  comment=[l]{\#},
  commentstyle=\color{purple}\ttfamily,
  stringstyle=\color{blue}\ttfamily,
  morestring=[b]',
  morestring=[b]"
}

\definecolor{addition}{rgb}{0, 0.6, 0} 
\definecolor{deletion}{rgb}{0.8, 0, 0} 
\definecolor{unchanged}{rgb}{0, 0, 0}  

\lstdefinelanguage{jac}{
       keywords={ import, node, ignore, take, activity, exit, spawn,
               edge, walker, and, or, if, elif, else, for, with, by, while, for, return,report, 
               continue, break, disengage, report, anchor, has, can, true, false, from, root, 
               context, info, details, try, strict, length, test, digraph, subgraph, test, by, in, to,
               skip, assert, yield, class, obj, Method, can, glob, from
           },
       sensitive=false, 
       commentstyle=\color{purple}\ttfamily,
       morecomment=[l]{//}, 
       morecomment=[l]{\#}, 
       morestring=[b]", 
       morestring=[b]',
    }
    
\newcommand{\pyinputlisting}[2][]{%
    \lstinputlisting[#1]{#2}%
}

\begin{document}

\title{Object-Spatial Programming}

\author{Jason Mars}
\affiliation{
  \institution{University of Michigan }
  \streetaddress{2260 Hayward St.}
  \city{Ann Arbor}
  \state{Mighigan}
  \country{USA}
  \postcode{48104}
}
\email{profmars@umich.edu}

\begin{abstract}
The evolution of programming languages from low-level assembly to high-level abstractions demonstrates a fundamental principle: by constraining how programmers express computation and enriching semantic information at the language level, we can make previously undecidable program properties tractable for optimization. Building on the insight of this \textbf{undecidability-lessening effect}, we introduce \textbf{Object-Spatial Programming (OSP)}, a novel programming model that extends Object-Oriented Programming by introducing topologically-aware class constructs called \textbf{archetypes}. OSP fundamentally inverts the traditional relationship between data and computation, enabling computation to move to data through four specialized archetypes: object classes, node classes (discrete data locations), edge classes (first-class relationships), and walker classes (mobile computational entities). By making topological relationships and traversal patterns explicit at the language level, OSP transforms previously opaque program behaviors into observable, optimizable patterns. This semantic enhancement enables runtime systems to make informed decisions about data locality, parallel execution, and distribution strategies based on explicit topology, while providing programmers with intuitive abstractions for modeling complex systems where connection topology is central to the computational model. The paradigm addresses fundamental limitations in traditional programming models when representing agent-based systems, social networks, neural networks, distributed systems, finite state machines, and other spatially-oriented computational problems, demonstrating how thoughtful abstraction design can simultaneously enhance programmer expressiveness and enable sophisticated system-level optimizations across the computing stack.
\end{abstract}

\maketitle

\section{Introduction}
\label{sec:introduction}

The evolution of programming languages from assembly to high-level abstractions represents more than mere convenience—it fundamentally alters what computing systems can understand and optimize about program behavior. At the heart of computer science theory lies the concept of undecidability, establishing fundamental limits on what can be algorithmically determined about programs. However, as we raise the level of abstraction in programming languages, we observe an important phenomenon: by constraining how programmers express computation and enriching the semantic information available at the language level, we can make previously undecidable properties tractable for specific, practical cases. This \textbf{undecidability-lessening effect} forms the foundational inspiration for a new class of programming abstractions that bridge the gap between program intent and system optimization.

While Object-Oriented Programming (OOP) advanced our ability to model entities and their behaviors, it reaches a representational limit when dealing with the relationships between objects and the patterns of interaction among them. In OOP, relationships are typically encoded as references or collections—semantically impoverished constructs that reveal little about the nature, purpose, or traversal patterns of these connections. This limitation becomes particularly pronounced in domains where the topology of connections is central to the computational model: social networks, agent-based systems, neural networks, distributed systems, and other topologically-oriented problems where computation logically flows through networks of interconnected entities.

Traditional programming models, including procedural, functional, and object-oriented paradigms, typically separate data structures from the algorithms that manipulate them. In this conventional model, data flows to computation through parameter passing and return values, with algorithms remaining stationary in functions and methods. This "data-to-compute" programming model is pervasive, being ubiquitous at the programming interface and fundamentally embedded in the von Neumann computer design itself, where data is moved from memory to the CPU for processing. Notably absent from the programming language landscape is a system of language constructs that naturally supports a complementary "compute-to-data" paradigm while maintaining compatibility with conventional programming languages.

The lack of such a programming model creates friction when representing computational problems where the topology of connections is central to the underlying computational model, where computation logically flows through a network of interconnected entities, where behavior is context-dependent based on tight couplings between data and compute, and where traversal patterns are complex and dynamically determined. Though graph algorithms and data structure libraries can be constructed in most programming languages, they remain secondary citizens, without first-class language support for topological semantics. This forces developers to implement complex traversal logic, maintain graph integrity, and manage event propagation through ad-hoc mechanisms that are often error-prone and difficult to maintain.

Beyond programmability advantages, embedding topological abstractions at the language level provides the runtime environment with rich semantic information about program behavior that would otherwise be obscured in conventional programming models. This heightened awareness of spatial relationships and traversal patterns enables a new class of optimizations particularly relevant to modern hardware architectures. The runtime can make informed decisions about data locality, parallel execution paths, and distribution strategies based on the explicit topology of the computation graph. Independent walker traversals can be automatically parallelized across processing cores, edge connections can inform predictive prefetching strategies, and the declarative nature of topological relationships enables automated reasoning about program correctness and performance characteristics.

To address these challenges and leverage the undecidability-lessening principle, we introduce \textbf{Object-Spatial Programming (OSP)}, a novel programming model that fundamentally inverts the relationship between data and computation. Rather than moving data to stationary computation units, OSP enables computation to move to data through topologically-aware constructs. This paradigm extends the semantics of Object-Oriented Programming by introducing specialized class-like constructs called \textbf{archetypes} that formalize spatial and topological relationships within the programming model itself.

At the core of Object-Spatial Programming are four key archetypes that extend traditional class semantics:

\begin{enumerate}
    \item \textbf{Object Classes} (\(\tau_{\text{obj}}\)): The universal supertype from which all other archetypes inherit, providing backward compatibility with traditional OOP concepts.
    
    \item \textbf{Node Classes} (\(\tau_{\text{node}}\)): Extensions of object classes that represent discrete locations or entities within a topological structure, capable of hosting computation and connecting to other nodes.
    
    \item \textbf{Edge Classes} (\(\tau_{\text{edge}}\)): First-class entities that represent relationships between nodes, encoding both the topology of connections and the semantics of those relationships. Edges serve not only as connections but also as potential locations where computation can reside.
    
    \item \textbf{Walker Classes} (\(\tau_{\text{walker}}\)): Autonomous computational entities that traverse the node-edge structure, residing and executing on both nodes and edges as they move through the topological space, carrying state and behaviors that execute based on their current location.
\end{enumerate}

Together, these archetypes create a complete topological representation framework where data (in nodes), relationships (as edges), and computational processes (through walkers) are explicitly modeled and integrated. This integration enables the paradigm shift from "data moving to computation" to "computation moving to data" while providing the semantic richness necessary for sophisticated system-level optimizations.

The semantic enhancement achieved through OSP exemplifies the undecidability-lessening principle in action. By making topological relationships and traversal patterns explicit at the language level, OSP transforms previously opaque program properties into observable, optimizable patterns. Edges as first-class relationships carry semantic meaning about the type, directionality, and computational significance of connections. Path collections explicitly capture traversal patterns that in OOP would be implicit in algorithm implementations. Walkers as mobile computation make explicit what would be hidden in call stacks and control flow, revealing patterns of locality, parallelism, and distribution that enable automated optimization across the computing stack.

The OSP paradigm offers significant advantages across numerous domains: for agent-based systems, walkers provide direct representation for autonomous agents that navigate environments; in distributed systems, the decoupling of data (nodes) from computation (walkers) creates natural models for distributed execution; in social networks and graph-based systems, it enables intuitive representations through natural mapping of users, relationships, and content to nodes and edges; and for finite state machines, states map naturally to nodes, transitions to edges, and execution flow to walker traversal. These examples represent just a few applications, as the paradigm's flexibility extends to any domain where connection topology is fundamental to the problem space.

This paper makes the following contributions:

\begin{enumerate}
    \item We establish the theoretical foundation of Object-Spatial Programming through the undecidability-lessening principle, demonstrating how thoughtful abstraction design can simultaneously enhance programmer expressiveness and enable sophisticated system-level optimizations.
    
    \item We formalize Object-Spatial Programming as an extension to Object-Oriented Programming, introducing four distinct archetypes: object classes, node classes, edge classes, and walker classes with their associated semantic constraints.
    
    \item We define a comprehensive semantic model that specifies how these archetypes interact, including instantiation rules, lifecycle management, and execution semantics for traversal operations within distributed computational contexts.
    
    \item We introduce specialized operators and statements for object-spatial execution, including the spawn operator (\(\Rightarrow\)) for activating computational entities and the visit statement (\(\triangleright\)) for traversing topological structures.
    
    \item We present the concept of \textbf{abilities} as a new function type with implicit execution semantics, triggered by spatial events rather than explicit invocation, enabling computation to be distributed throughout the topological structure.
    
    \item We demonstrate the practical application of OSP through a detailed case study of a social media application that naturally maps domain concepts to object-spatial constructs, illustrating how the paradigm simplifies complex traversal algorithms.
\end{enumerate}

The remainder of this paper is organized as follows: Section~\ref{sec:decidability-insight} establishes the theoretical foundation through the undecidability-lessening insight and its implications for programming language design. Section~\ref{sec:semantics} formalizes the semantic elements of the OSP model, including archetypes, execution semantics, and ability definitions. Section~\ref{sec:jac-implementation} presents a detailed case study of a social media application implemented in Jac, a language that embodies the OSP paradigm. We conclude with a discussion of implementation considerations and future directions for research in this emerging programming paradigm that bridges theoretical computer science insights with practical system optimization opportunities.

\section{The Undecidability-Lessening Insight}
\label{sec:decidability-insight}

The creation of Object-Spatial Programming stems from a fundamental insight that bridges the theoretical foundations of computer science with its most practical implementations. This insight recognizes that the evolution of programming languages from low-level assembly to high-level abstractions represents more than mere convenience—it fundamentally alters what computing systems can understand and optimize about program behavior.

\subsection{The Undecidability-Lessening Effect}

At the heart of computer science theory lies the concept of undecidability—the fundamental limits on what can be algorithmically determined about program behavior. The halting problem and Rice's theorem establish that most non-trivial semantic properties of programs are undecidable in the general case. However, as we raise the level of abstraction in programming languages, we observe an important phenomenon: by constraining how programmers express computation and enriching the semantic information available at the language level, we can make previously undecidable properties tractable for specific, practical cases.

When programmers write in assembly language, the gap between the program's syntactic representation and its semantic intent is vast. The system sees only sequences of low-level operations with minimal context about their purpose. As we ascend through procedural languages, object-oriented programming, and functional paradigms, each abstraction layer adds semantic constraints and contextual information that narrow this gap. The programming interface becomes progressively richer and closer to expressing the true intent of the program.

\subsection{Beyond Objects: The Semantic Richness of Topology}

While Object-Oriented Programming (OOP) advanced our ability to model entities and their behaviors, it reaches a representational limit when dealing with the relationships between objects and the patterns of interaction among them. In OOP, relationships are typically encoded as references or collections—semantically impoverished constructs that reveal little about the nature, purpose, or traversal patterns of these connections. The introduction of topologically-structured computation in OSP represents a key elevation in abstraction level that captures previously inexpressible program semantics.

Consider the semantic information that emerges when we move from OOP to OSP:

\begin{itemize}
    \item \textbf{Edges as First-Class Relationships}: Rather than mere references, edges in OSP carry semantic meaning about the type, directionality, and computational significance of relationships. An edge between a User node and a Post node can encode not just that a relationship exists, but what kind of relationship it represents, what computations should occur during traversal, and how that relationship participates in larger patterns.
    
    \item \textbf{Paths as Computational Intent}: Path collections explicitly capture traversal patterns that in OOP would be implicit in algorithm implementations. A path through a social graph from users to their friends' recent posts is not just a sequence of method calls but a declared route through the data topology that reveals the algorithm's intent and access patterns.
    
    \item \textbf{Walkers as Mobile Computation}: The movement of computation through data via walkers makes explicit what in OOP would be hidden in call stacks and control flow. The system can observe not just that data is being accessed, but how computational processes flow through the data topology, revealing patterns of locality, parallelism, and distribution.
\end{itemize}

This topological enrichment provides programming interfaces that allow developers to articulate their programs' behavior with higher-level semantic clarity. The relationships between entities, the patterns of traversal through those relationships, and the flow of computation across the data landscape all become explicit, analyzable properties of the program rather than emergent behaviors hidden in implementation details.

\subsubsection{Illustrative Example}

To illustrate this limitation of OOP and the power of topological abstractions, consider the classic Vehicle hierarchy—a canonical example of OOP design. We can elegantly model Cars, Trucks, and Motorcycles as classes inheriting from Vehicle, with well-defined properties like engine, wheels, and methods like accelerate() or brake(). This abstraction excels at representing the components and behaviors of individual vehicles.

However, this same model breaks down when we need to express the relationships between vehicles and the patterns of their interactions. How do we model the relationship between vehicles in traffic? A Car following another Car is fundamentally different from a Car following a Motorcycle (different safe distances, different visibility concerns). The relationship between an emergency vehicle and other vehicles carries special semantics (must yield, pull over). These relationships in OOP are typically reduced to mere references or collections—a Car might have a list of "nearbyVehicles" or a reference to "vehicleAhead," but these constructs reveal nothing about the nature, rules, or dynamics of these relationships in a representation that can be analyzed by underlying systems.

In a OSP model of traffic, these relationships become first-class citizens with their own semantics and behaviors. A "Following" edge between two Vehicle nodes is not merely a reference—it's a first-class entity that can encode following distance, relative speed, safety constraints, and regulatory rules. An "Emergency Yielding" edge carries different semantics than a "Normal Following" edge. Paths can represent concepts like "all vehicles that must yield to an approaching ambulance," making routing and right-of-way algorithms explicit and analyzable. Walkers like a "Route Optimizer" can traverse through the traffic network, carrying optimization algorithms to the data rather than pulling all vehicle data to a central algorithm. This transformation from implicit relationships to explicit topological constructs exemplifies how OSP enables clearer expression of program intent while providing richer information for system optimization.

\subsection{Empowering Runtime and Compilation Systems}

This enrichment of semantic information through topological abstractions has profound practical implications. Compilation and runtime systems gain access to higher-level program properties that were previously opaque. With more information about program intent encoded directly in the language constructs, these systems can:

\begin{itemize}
    \item Automate complex mappings from high-level intent to low-level system interfaces
    \item Hide architectural complexities from programmers while maintaining or improving performance
    \item Make informed decisions about resource allocation, parallelization, and distribution based on explicit topological patterns
    \item Perform optimizations that would be impossible without semantic awareness of relationship structures and traversal patterns
    \item Predict and prefetch data based on declared paths rather than attempting to infer access patterns
\end{itemize}

The relationship is bidirectional: as programming models become more expressive and semantically rich, they simultaneously enhance the programmer's ability to articulate intent and provide underlying systems with a symbolic representation that improves the decidability of that intent. This creates a virtuous cycle where higher abstraction enables better optimization, which in turn makes higher abstraction more practical.

\subsubsection{Concrete Examples}
The semantic richness of OSP enables new classes of optimization opportunities that span from low-level hardware management to high-level distributed system coordination. By making topological relationships and walker traversal patterns explicit at the language level, OSP provides compilers and runtime systems with new visibility into program behavior, enabling automated optimizations that would be impossible with traditional programming models.

\begin{table}[ht]
\centering\tiny
\begin{tabular}{|p{3cm}|p{8cm}|}
\hline
\textbf{Optimization Domain} & \textbf{OSP-Enabled Capabilities} \\
\hline
\textbf{Runtime Database Schema Design and Adaptation} & The explicit node-edge topology directly informs automated database schema generation, where nodes map to tables and edges to foreign key relationships. Compiler analysis of walker traversal patterns reveals query access patterns, enabling automatic index creation, table partitioning strategies, and materialized view generation. Path collections translate directly to query optimization hints, while walker locality patterns inform distributed database sharding decisions and replica placement strategies. \\
\hline
\textbf{Processing-in-Memory} & Node-bound abilities map naturally to processing-in-memory architectures, where computation occurs directly at data storage locations. Walker arrivals trigger localized computation without data movement, while edge traversals can exploit near-data processing capabilities. The runtime can automatically partition computation between traditional processors and processing-in-memory units based on data locality patterns revealed by walker traversal analysis. \\
\hline
\textbf{Compilation-Time Analysis} & Unlike traditional programs where data access patterns emerge only at runtime, OSP's explicit topology enables powerful static analysis. Compilers can detect potential race conditions in concurrent walker traversals, optimize memory layouts based on path connectivity, and generate specialized code paths for different walker-node type combinations. The declarative nature of abilities allows for ahead-of-time compilation of location-specific computational kernels. \\
\hline
\textbf{Accelerator Hardware Integration} & GPU and TPU optimization becomes tractable when walker spawning patterns reveal data parallelism opportunities. Independent walker trajectories can be automatically distributed across SIMD units, while the explicit node-edge structure enables efficient memory coalescing. Walker locality patterns inform optimal data placement on accelerator memory hierarchies. \\
\hline
\textbf{Scale-Agnostic Application Development} & OSP's topological abstractions enable scale-agnostic programming where applications are written once and executed across dramatically different deployment contexts without modification. The explicit topology provides sufficient semantic information for runtime systems to automatically handle persistence mapping, multi-user context isolation, API generation, and distributed execution. Compilers can generate deployment-specific optimizations—single-user applications might use in-memory data structures, while multi-user deployments automatically gain transactional consistency and concurrent access control. \\
\hline
\textbf{Cloud and Container Orchestration} & The separation of stationary data (nodes) from mobile computation (walkers) creates natural boundaries for container scaling and distribution. Runtime systems can automatically spawn containers near frequently accessed node clusters, while walker migration patterns inform load balancing decisions. Path collections enable predictive resource allocation, allowing cloud systems to pre-provision computational resources along anticipated walker routes. \\
\hline
\end{tabular}
\caption{OSP-Enabled Optimization Opportunities Across Computing Domains}
\label{tab:dsp-optimizations}
\end{table}

The optimization opportunities show in Table~\ref{tab:dsp-optimizations} represent a few examples that emerge directly from OSP's elevation of topological semantics to first-class language constructs, demonstrating how thoughtful abstraction design can simultaneously enhance programmer expressiveness and enable sophisticated system-level optimizations across the entire computing stack.

\subsection{Object-Spatial Programming as Semantic Enhancement}

Object-Spatial Programming represents a deliberate application of this insight to the domain of topologically-structured computation. By introducing archetypes (nodes, edges, walkers) and specialized operators (spawn, visit) as first-class language constructs, OSP creates a programming model where:

\begin{enumerate}
    \item \textbf{Topological Intent is Explicit}: Rather than encoding graph structures and traversal patterns in ad-hoc data structures and algorithms, OSP makes these concepts part of the language itself. The runtime system can directly observe and reason about the topological relationships in the program.
    
    \item \textbf{Computational Flow is Declarative}: The movement of walkers through the node-edge structure provides a clear, observable model of how computation flows through data. This makes previously implicit patterns of data access and computation scheduling explicit and analyzable.
    
    \item \textbf{Distribution Boundaries are Semantic}: By separating data (nodes), relationships (edges), and computation (walkers), OSP provides natural boundaries for distribution and parallelization that align with program semantics rather than arbitrary implementation choices.
\end{enumerate}

\subsection{A Design Principle for Future Systems}

A critical corollary of this insight is that \textbf{systems and runtime environments should always be designed to exploit language-level abstractions}. The history of computing demonstrates that when system designers ignore the semantic information provided by higher-level abstractions, they miss optimization opportunities. Just as garbage collectors exploit type information, JIT compilers exploit method invocation patterns, and GPUs exploit data parallelism expressed in kernel languages, future systems must be designed with deep awareness of the abstractions they support.

For OSP, this means runtime systems should:
\begin{itemize}
    \item Analyze edge types and walker patterns to optimize data layout
    \item Exploit path declarations for intelligent prefetching and caching
    \item Recognize independent walker trajectories for automatic parallelization
    \item Use topological information for efficient distribution strategies
\end{itemize}

This design principle extends beyond OSP to any future programming model: the abstractions provided by a language are not merely conveniences for programmers but rich sources of semantic information that system designers must actively exploit to achieve optimal performance.

The key insight is that by raising the level of abstraction to include topological and traversal semantics as first-class citizens, OSP transforms previously undecidable or difficult-to-analyze properties into observable, optimizable patterns. This exemplifies how thoughtful language design can simultaneously simplify programming and enable sophisticated system-level optimizations, fulfilling the dual promise of enhancing programmer expressiveness while providing rich semantic information to underlying systems.
\section{Object-Spatial Topological Semantics}
\label{sec:semantics}
The foundational concept of Object-Spatial Programming is the formalization of topological relationships through the introduction of special class types and a handful of new language constructs. This section outlines the core semantic elements of these constructs related to object-spatial topology, which fundamentally inverts the traditional relationship between data and computation.

\subsection{Unified Notation}
\label{subsec:notation}

To ensure clarity and consistency throughout the formalization of Object-Spatial Programming, we define the following notation:

\begin{table}[ht]
\centering\tiny
\begin{tabular}{|l|l|}
\hline
\textbf{Symbol} & \textbf{Definition} \\
\hline
\multicolumn{2}{|c|}{\textbf{Classes and Instances}} \\
\hline
$C$ & Set of all class definitions in the programming model \\
$\tau_{\text{obj}}$ & Object class type (universal supertype) \\
$\tau_{\text{node}}$ & Node class type (extends object classes) \\
$\tau_{\text{edge}}$ & Edge class type (extends object classes) \\
$\tau_{\text{walker}}$ & Walker class type (extends object classes) \\
$n, n_i, n_{\text{src}}, n_{\text{dst}}$ & Instances of node classes \\
$e, e_i$ & Instances of edge classes \\
$w, w'$ & Instances of walker classes \\
$o$ & Generic instance of any class \\
\hline
\multicolumn{2}{|c|}{\textbf{Object-Spatial Constructs}} \\
\hline
$\mathcal{P}$ & Path collection (ordered collection of nodes or edges) \\
$\mathcal{P}_N$ & Node path (ordered collection of connected nodes) \\
$\mathcal{P}_E$ & Edge path (ordered collection of connected edges) \\
$Q_w$ & Walker's traversal queue \\
\hline
\multicolumn{2}{|c|}{\textbf{Operations and References}} \\
\hline
$\Rightarrow$ & Spawn operator (activates a walker) \\
$\triangleright$ & Visit statement (adds to walker's traversal queue) \\
$a_{\text{walker}}$ & Walker ability (triggered during traversal) \\
$a_{\text{node}}$ & Node ability (triggered by walker visits) \\
$a_{\text{edge}}$ & Edge ability (triggered by walker traversal) \\
$a^{\text{entry}}$ & Entry ability (triggered on arrival) \\
$a^{\text{exit}}$ & Exit ability (triggered on departure) \\
$\mathbf{self}$ & Self-reference within an instance \\
$\mathbf{here}$ & Contextual reference to current walker's edge or node location \\
$\mathbf{visitor}$ & Contextual reference to current location's visiting walker \\
$\mathbf{path}$ & Contextual reference to current walker's destination queue $Q_w$ \\
$\prec$ & Execution precedence relation between operations; $a \prec b$ means 
         operation $a$ must complete before operation $b$ begins \\
\hline
\multicolumn{2}{|c|}{\textbf{Set and Logical Notation}} \\
\hline
$x \in X$ & $x$ is an element of set $X$ \\
$A \subseteq B$ & Set $A$ is a subset of set $B$ \\
$A \cup B$ & Union of sets $A$ and $B$ \\
$A \cap B$ & Intersection of sets $A$ and $B$ \\
$\wedge$ & Logical AND \\
$\vee$ & Logical OR \\
$\neg$ & Logical NOT \\
$\forall$ & Universal quantification ("for all") \\
$\exists$ & Existential quantification ("there exists") \\
$f : X \rightarrow Y$ & Function $f$ mapping from domain $X$ to codomain $Y$ \\
$:=$ & Definition or assignment \\
$=$ & Equality test \\
\hline
\end{tabular}
\caption{Notation used in the OSP formalization}
\label{tab:notation}
\end{table}

\subsection{Archetypes of Classes}

We define four distinct \textbf{archetype} classes, extending the traditional class paradigm to incorporate object-spatial semantics:

\begin{enumerate}
    \item \textbf{Object Classes} (\(\tau_{\text{obj}}\)): These are conventional classes, analogous to traditional OOP class types. Objects can have properties that describe their intrinsic characteristics and methods that operate on those properties. They serve as the foundational building blocks from which other archetypes derive, maintaining backward compatibility with existing OOP concepts while enabling integration with object-spatial extensions.
    
    \item \textbf{Node Classes} (\(\tau_{\text{node}}\)): These extend object classes and can be connected via edges. Nodes represent discrete locations or entities within a topological graph structure. They encapsulate data, compute, and the potential for connections, serving as anchoring points in the object-spatial topology of the program. In addtion to object semantics, nodes bind computation to data locations through \textit{abilities}, allowing execution to be triggered by visitation rather than explicit invocation.
    
    \item \textbf{Edge Classes} (\(\tau_{\text{edge}}\)): These represent directed relationships between two node instances and can only be instantiated when two nodes are specified. Edges encode both the topology of connections and the semantics of those connections. Unlike simple references in traditional OOP, edges are first-class object entities with their own properties and behaviors, enabling rich modeling of connection types, weights, capacities, or other relationship attributes. Importantly, edges serve not only as connections but also as traversable locations for walkers, with their own computational context.
    
    \item \textbf{Walker Classes} (\(\tau_{\text{walker}}\)): These model autonomous entities that traverse node and edge objects. Walkers represent active computational elements that move through the data topological structure, processing data or triggering behaviors as they visit different nodes and edges. They enable decoupling of traversal logic from data structure, allowing for modularity in algorithm design and implementation. Walkers embody the paradigm shift of OSP, carrying computational behaviors to data rather than data being passed to computation.
\end{enumerate}

This archetype system creates a complete topological representation framework, where data (in nodes and edges), relationships (as edges), and computational processes (through walkers) are all explicitly modeled and integrated, inverting the traditional paradigm of passing data to functions.

\subsubsection{Formalization}
\label{subsubsec:formalization}

Let \( C \) be the set of all class definitions in the programming model, where:

\begin{enumerate}
    \item \( \tau_{\text{obj}} \in C \) is a standard object class type, representing the universal supertype from which all other archetypes inherit.
    
    \item \( \tau_{\text{node}} \subseteq \tau_{\text{obj}} \) represents node class types, which extend object classes with connectivity capabilities and data-bound computation. This subset relationship ensures that nodes inherit all capabilities of objects while adding topological semantics and the ability to bind computation to data locations.
    
    \item \( \tau_{\text{edge}} \subseteq \tau_{\text{obj}} \) represents edge class types, which extend object classes with relational semantics. Edges are not merely references but full-fledged objects that encapsulate relationship properties and behaviors, serving as both connections and traversable locations.
    
    \item \( \tau_{\text{walker}} \subseteq \tau_{\text{obj}} \) represents walker class types, which extend object classes with mobility semantics within the node-edge structure. Walkers combine data, state, and traversal logic to model computational processes that flow through the topological structure, actualizing the concept of "computation moving to data."
\end{enumerate}

Each instance \( n \) of a \textbf{node class} \( \tau_{\text{node}} \) is defined simply as: \( n = () \).
Nodes exist as independent entities in the topological structure, serving as primary data locations. Unlike edges which require references to nodes, nodes can exist without connections to other elements, though they typically participate in the graph structure through edges that reference them.

Each instance \( e \) of an \textbf{edge class} \( \tau_{\text{edge}} \) is defined as a tuple:

\[
e = (n_{\text{src}}, n_{\text{dst}})
\]

where:

\begin{itemize}
    \item \( n_{\text{src}}, n_{\text{dst}} \in \tau_{\text{node}} \) are the source and destination node instances, serving as the endpoints of the relationship. These must exist prior to edge creation, establishing a dependency constraint that maintains object-spatial graph integrity.
\end{itemize}

This formalization ensures that edges properly connect existing nodes, maintaining topological graph consistency within the program.

Each instance \( w \) of a \textbf{walker class} \( \tau_{\text{walker}} \) may exist in one of three states:

\[
w = \begin{cases}
(n_{\text{loc}}) & \text{if active within node topological context} \\
(e_{\text{loc}}) & \text{if active within edge topological context} \\
() & \text{if inactive as a standard object}
\end{cases}
\]

where \( n_{\text{loc}} \in \tau_{\text{node}} \) is the node the walker resides on, or \( e_{\text{loc}} \in \tau_{\text{edge}} \) is the edge the walker traverses when active. This location property is dynamic and changes as the walker traverses the topological structure, allowing the walker to access different data contexts based on its current position. When inactive, the walker exists as a standard object without object-spatial context, allowing for manipulation before activation within a object-spatial context.

The \textbf{overall topological structure} with active computational elements can be represented as:

\[
G = (N, E, W, L)
\]

where:
\begin{itemize}
    \item \( N \) is the set of all node instances, representing data locations in the topological space
    \item \( E \) is the set of all edge instances, defining the connectivity between nodes
    \item \( W \) is the set of all walker instances, representing the computational entities
    \item \( L: W \rightarrow N \cup E \cup \{\emptyset\} \) is a location mapping function that associates each walker with its current position in the topology, where \(\emptyset\) indicates an inactive walker
\end{itemize}

This representation captures both the static structural components (nodes and edges) and the dynamic computational elements (walkers) of the object-spatial system, along with their current positions within the topological structure. Next, we introduce the first-class construct of a path collection which represents a traversable path through a object-spatial topology.

\subsection{Path Collections (\(\mathcal{P}\)) and Walker Destination Queues (\(Q_w\))}
\label{subsubsec:pathcollection}

The Object-Spatial Programming model introduces two complementary constructs that govern traversal dynamics: \textbf{path collections} (\(\mathcal{P}\)), which represent potential traversal routes through the topology, and \textbf{walker destination queues} (\(Q_w\)), which manage the actual execution sequence during traversal. Together, these constructs create a flexible yet deterministic framework for computational movement through data spaces.

\subsubsection{Path Collections}
The \textbf{path collection} (\(\mathcal{P}\)) introduces a higher-order topological construct that represents an ordered sequence of nodes and edges within the object-spatial structure. As first-class citizens in the programming model, path collections can be created, modified, and manipulated like any other data structure. This abstraction serves as a critical link between topology and traversal semantics, enabling concise expression of traversal patterns while maintaining the integrity of the object-spatial model.

The intent of the path collection is to provide a unified framework that bridges graph theory and computation, creating a formal way to express how walkers move through connected data structures. Rather than treating node traversals and edge traversals as separate concerns, the path collection unifies them into a single construct that preserves topological relationships while enabling richer expression of traversal algorithms.

A path collection is defined as:
\[
\mathcal{P} = [p_1, p_2, \ldots, p_k]
\]
where each \(p_i \in N \cup E\) (i.e., each element is either a node or an edge), subject to the following constraints:
\begin{enumerate}
    \item \textbf{Origin Connectivity:} The first element $p_1$ must be connected to an origin node $n_{\text{origin}} \in N$, either by being the origin itself or by being an edge with $n_{\text{origin}}$ as an endpoint.
    
    \item \textbf{Sequential Connectivity:} For each element $p_i$ where $i > 1$ in $\mathcal{P}$, at least one of the following must hold:
    \begin{itemize}
        \item If $p_i$ is a node, then there must exist at least one element $p_j$ where $j < i$ such that either:
        \begin{itemize}
            \item $p_j$ is a node and there exists an edge $e \in E$ connecting $p_j$ and $p_i$, or
            \item $p_j$ is an edge with $p_i$ as one of its endpoints
        \end{itemize}
        
        \item If $p_i$ is an edge, then at least one of its endpoints must appear as a node in $\{p_1, p_2, \ldots, p_{i-1}\}$
    \end{itemize}
    
    \item \textbf{Path Completeness:} For any element $p_i$ in $\mathcal{P}$, there must exist a path from $n_{\text{origin}}$ to $p_i$ such that all intermediate elements (nodes and edges) on that path are present in the prefix $\{p_1, p_2, \ldots, p_{i-1}\}$. This ensures that the path collection contains at least one valid traversal route to each included element.
    
    \item \textbf{Traversal Coherence:} When multiple elements are eligible for inclusion at a given point in the sequence (e.g., multiple nodes connected to previously included elements), their relative ordering follows breadth-first search (BFS) semantics from the most recently added elements, preserving locality of traversal.
\end{enumerate}

This definition ensures topological validity by anchoring all nodes to a common origin while allowing flexible expression of traversal patterns. The breadth-first ordering provides predictability and consistency for walkers traversing the path, particularly when dealing with hierarchical structures where multiple branches might need to be explored.

As first-class citizens, path collections support arbitrary modifications, including additions, removals, reorderings, and transformations, provided that the resulting collection maintains the properties of a valid path collection. Operations such as concatenation, slicing, filtering, and mapping can be applied to path collections, yielding new valid path collections. This flexibility enables algorithmic manipulation of potential traversal paths while preserving the topological integrity of the underlying data structure.

This generalized path collection model reflects a natural way to describe, in a declarative way, potential routes as to how walker may navigate through a data topology. By allowing both nodes and edges in the same sequence while maintaining topological context, path collections enable algorithms to be expressed in terms of the connected data structures they operate on, rather than as abstract operations that receive data as input.

\subsubsection{Path Construction}
Building on the definition of path collections as first-class citizens in the programming model, a path collection can be constructed in several ways:

\begin{enumerate}
    \item \textbf{Explicit Construction}: By directly specifying the ordered sequence of nodes and optional edges:
    
    \[
    \mathcal{P} = [p_1, p_2, \ldots, p_k] \text{ where } p_i \in N \cup E
    \]
    
    When explicitly constructing path collections, the elements must satisfy the reachability and ordering constraints defined earlier, ensuring that walkers can traverse through the path in a topologically valid sequence. Edges, when included, must immediately precede the nodes they connect to on the reachability path from the origin node.
    
    \item \textbf{Query-Based Construction}: By specifying an origin node and a traversal predicate:
    
    \[
    \mathcal{P} = \text{path}(n_{\text{origin}}, \text{predicate}, \text{includeEdges}, d)
    \]
    
    where:
    \begin{itemize}
        \item \(n_{\text{origin}} \in N\) is the origin node from which all nodes in the path must be reachable
        \item \(\text{predicate}: N \cup E \rightarrow \{\text{true}, \text{false}\}\) is a function that determines whether an element should be included in the path
        \item \(\text{includeEdges} \in \{\text{true}, \text{false}\}\) specifies whether edges should be explicitly included in the path collection
        \item \(d \in \{\text{outgoing}, \text{incoming}, \text{any}\}\) specifies the traversal direction for constructing the path
    \end{itemize}
    
    The construction algorithm performs a breadth-first traversal from the origin node, adding elements to the path collection according to the predicate and the includeEdges parameter. This ensures that the resulting path maintains proper reachability relations while allowing flexible filtering of elements.
    
    A common example of query-based construction is creating a path that follows a specific sequence of edge types:
        
    \[
    \mathcal{P} = \text{path}(n_{\text{origin}}, \lambda e : e \in E \land \text{type}(e) \in [\tau_{\text{edge}}^1, \tau_{\text{edge}}^2, \ldots, \tau_{\text{edge}}^k] \text{ in sequence}, \text{true}, \text{outgoing})
    \]
        
    This creates a path collection starting at \(n_{\text{origin}}\) and following only edges that match the specified sequence of edge types. The path resolution algorithm traverses the graph, selecting edges and their connected nodes that conform to this type pattern. A walker traversing this path would follow a route determined by these edge type constraints, enabling declarative specification of complex traversal patterns based on relationship types.
\end{enumerate}

Once constructed, these path collections can be passed to walkers to guide their traversal through the topological structure, as we'll see in the next section on walker destination queues.

\subsubsection{Walker Destination Queues}
While path collections define potential traversal routes through the topology, \textbf{walker destination queues} (\(Q_w\)) represent the actual execution sequence that a walker follows during its traversal. Each active walker \(w\) maintains an internal traversal queue \(Q_w\) that determines its next destinations:
\[
Q_w = [q_1, q_2, \ldots, q_m] \text{ where } q_i \in N \cup E
\]
Walker destination queues have several key properties that govern traversal dynamics:
\begin{enumerate}
    \item \textbf{First-In-First-Out (FIFO) Processing}: Walker destination queues follow FIFO semantics, with elements processed in the order they were added. When a walker completes execution at its current location, it automatically moves to the next element in its queue.
    
    \item \textbf{Dynamic Modification}: Walker destination queues are designed for dynamic modification during traversal through visit statements (described in Section~\ref{subsubsec:visit}) and other control flow mechanisms:
    
    \[
    \text{visit}(w, n) \Rightarrow Q_w \leftarrow Q_w \cup [n]
    \]
    
    This allows walkers to adapt their traversal paths based on discovered data or computed conditions.
    
    \item \textbf{Automatic Edge-to-Node Transitions}: When a walker traverses an edge, the appropriate destination node is automatically added to its queue if not already present, ensuring continuity in the traversal process.
    
    \item \textbf{Path-to-Queue Conversion}: When a walker spawns on or visits a path collection, the path is converted into queue entries according to traversal requirements:
    
    \[
    \text{visit}(w, \mathcal{P}) \Rightarrow Q_w \leftarrow Q_w \cup \text{expandPath}(\mathcal{P}, L(w))
    \]
    
    where \(\text{expandPath}(\mathcal{P}, L(w))\) transforms the path collection into a physically traversable sequence from the walker's current location \(L(w)\). This function ensures that:
    
    \begin{itemize}
        \item All elements in \(\mathcal{P}\) are included in the expanded queue
        \item Any necessary intermediate nodes or edges required for physical traversal between non-adjacent elements are inserted
        \item The resulting sequence maintains the relative ordering of elements in the original path collection
        \item The expanded path respects the connectivity constraints of the topological structure
    \end{itemize}
    
    This conversion enables walkers to traverse path collections that express higher-level traversal intent without requiring explicit specification of every intermediate step.
    
    \item \textbf{Activity Persistence}: Once a walker transitions to an active state via spawn, it remains active until its queue is exhausted or it is explicitly disengaged. This ensures computational continuity during traversal, maintaining the walker's contextual state throughout its path exploration. When a walker's queue becomes empty after all abilities at its current location have executed, it automatically transitions back to an inactive state. However, while at a node with an empty queue, it temporarily preserves its active status, allowing for potential reactivation through new visit statements before the current execution cycle completes.
\end{enumerate}
The relationship between the path collection (\(\mathcal{P}\)) and the dynamic walker queue (\(Q_w\)) creates a flexible yet deterministic traversal model, allowing for both declarative path specifications and runtime adaptation of traversal behavior.

\subsection{Abilities}
\label{subsec:abilities}

In addition to traditional methods \( m: \tau \rightarrow \tau' \), we introduce \textbf{abilities}, a new function type \( a: \varnothing \rightarrow \varnothing \) with \emph{implicit} execution semantics. Unlike ordinary functions, abilities neither accept explicit arguments nor return values; instead, they gain access to relevant data through the walker or location (node or edge) that triggers them. This represents a fundamental paradigm shift: rather than moving data to computation through parameters and return values, computation is distributed throughout the topology and automatically activated by object-spatial interactions. Abilities are named using the same conventions as methods, providing a consistent interface pattern across the programming model.

Each ability specifies an execution trigger that determines when it is activated during traversal:

\paragraph{\textbf{Walker Abilities}} \( a_{\text{walker}} \) are automatically triggered when a walker enters or exits a node or edge of a specified type:
    \[
    a_{\text{walker}} : (\tau_{\text{location}}, t) \rightarrow \bot
    \]
    where $\tau_{\text{location}} \in \{\tau_{\text{node}}, \tau_{\text{edge}}\}$ and $t \in \{\text{entry}, \text{exit}\}$ specifies whether the ability is triggered upon the walker's entry to or exit from a location of the specified type. This notation indicates the \emph{condition} under which the ability is invoked, rather than a parameter list. The ability thus acts as an event handler for location arrival or departure events. Once invoked, the ability can directly access the triggering walker's data (via \(\mathbf{self}\)), the location it arrived at or is departing from (via \(\mathbf{here}\)), and its traversal path and queue (via \(\mathbf{path}\)). The walker can modify its traversal queue \(Q_w\) through visit statements or direct manipulation of \(\mathbf{path}\). This allows walkers to respond contextually to different location types they encounter, implementing type-specific processing logic without explicit conditional branching and dynamically adapting their traversal path based on discovered data. The walker serves both as a carrier of computational behavior and as an activator of location-bound computation throughout the topology.

\paragraph{\textbf{Node Abilities}} \( a_{\text{node}} \) are automatically triggered when a walker of a specified type enters or exits the node:
    \[
    a_{\text{node}} : (\tau_{\text{walker}}, t) \rightarrow \bot
    \]
    where $t \in \{\text{entry}, \text{exit}\}$ specifies whether the ability is triggered upon the walker's entry to or exit from the node. Similarly, this indicates the condition (arrival or departure of a walker of type \(\tau_{\text{walker}}\)), not an explicit parameter. The ability functions as an event handler for walker arrival or departure events. When triggered, the ability can access the node's data (via \(\mathbf{self}\)), the incoming or outgoing walker (via \(\mathbf{visitor}\)), and the walker's destination queue (via \(\mathbf{path}\)). This allows nodes to respond differently to different types of walkers, implementing specialized processing logic based on the visitor type and traversal stage, and potentially influencing the walker's future traversal path. Node abilities demonstrate that nodes are not merely passive data containers but active computational sites that respond to traversal events, embodying the distributed nature of computation in the OSP model.
    
\paragraph{\textbf{Edge Abilities}} \( a_{\text{edge}} \) are automatically triggered when a walker of a specified type enters or exits the edge:
    \[
    a_{\text{edge}} : (\tau_{\text{walker}}, t) \rightarrow \bot
    \]
    where $t \in \{\text{entry}, \text{exit}\}$ specifies whether the ability is triggered upon the walker's entry to or exit from the edge. This ability functions similarly to node abilities but is specific to edge contexts. When triggered, the ability can access the edge's data (via \(\mathbf{self}\)), the traversing walker (via \(\mathbf{visitor}\)), and the walker's destination queue (via \(\mathbf{path}\)). Edge abilities enable computational behavior to be bound to relationship transitions, allowing for processing that specifically occurs during the movement between nodes, including the possibility of modifying the walker's future traversal path. This enables modeling of transition-specific computation, such as filtering, transformation, or validation of data as it flows through the topological structure. The presence of computational abilities in edges reinforces that in the OSP model, even transitions between data locations are first-class citizens capable of containing and executing computation.

\subsubsection{Ability Execution Order}
\label{subsubsec:abilityexec}

When a walker traverses the topological structure, abilities are executed in a consistent, predictable order regardless of location type. This execution order respects both entry/exit specifications and the dual-perspective model of location-walker interaction:

\begin{enumerate}
    \item \textbf{Arrival Phase} - When a walker arrives at any location (node or edge):
    \begin{enumerate}
        \item First, all relevant location entry abilities for the arriving walker type are executed. This allows the location (node or edge) to respond to the walker's arrival, potentially modifying its own state or the state of the walker. This represents location-bound computation triggered by mobile traversal.
        
        \item Next, all relevant walker entry abilities for the current location type are executed. This allows the walker to respond to its new context, potentially modifying its own state or the state of the location. This represents mobile computation processing data at its current position.
        
        \item During execution of these abilities, the walker may modify its traversal queue \(Q_w\) through visit statements if at a node, or have its queue automatically updated if at an edge:
        \begin{itemize}
            \item At nodes: The walker may execute visit statements to enqueue new destinations
            \item At edges: The appropriate endpoint node is automatically added to \(Q_w\) if not already present, based on traversal direction:
            \begin{itemize}
                \item If arrived from \(n_{\text{src}}\), then \(Q_w \leftarrow Q_w \cup [n_{\text{dst}}]\)
                \item If arrived from \(n_{\text{dst}}\), then \(Q_w \leftarrow Q_w \cup [n_{\text{src}}]\)
            \end{itemize}
        \end{itemize}
    \end{enumerate}

    \item \textbf{Departure Phase} - When a walker prepares to leave any location (node or edge):
    \begin{enumerate}
        \item First, all relevant walker exit abilities for the current location type are executed. This allows the walker to finalize any processing before departure.
        
        \item Next, all relevant location exit abilities for the departing walker type are executed. This allows the location to respond to the walker's departure, potentially performing cleanup or transition operations.
        
        \item After all exit abilities have executed, the walker updates its location to the next element in its queue: \(L(w) \leftarrow \text{dequeue}(Q_w)\)
    \end{enumerate}
    
    \item \textbf{Queue Exhaustion} - When a walker's queue \(Q_w\) becomes empty after dequeuing:
        \begin{enumerate}
        \item If the walker is on a node, it remains at that node until further visit statements are executed or until explicitly disengaged
        
        \item If the walker is on an edge, the program raises an error, as edges cannot be terminal locations
    \end{enumerate}
\end{enumerate}

This unified execution order establishes a predictable pattern where locations first respond to walker arrival before walkers process their new context, and walkers prepare for departure before locations respond to their exit. The order is expressed formally as:

\[
\begin{array}{c}
\forall a_{\text{loc}}^{\text{entry}} \in l, \forall a_{\text{walker}}^{\text{entry}} \in w : \text{execute}(a_{\text{loc}}^{\text{entry}}) \prec \text{execute}(a_{\text{walker}}^{\text{entry}}) \\
\forall a_{\text{walker}}^{\text{exit}} \in w, \forall a_{\text{loc}}^{\text{exit}} \in l : \text{execute}(a_{\text{walker}}^{\text{exit}}) \prec \text{execute}(a_{\text{loc}}^{\text{exit}})
\end{array}
\]

Additionally, the relationship between queue operations and ability execution follows this pattern:

\[
\begin{array}{c}
\forall w \in W, \forall l \in L(w) : \text{execute-all-abilities}(w, l) \prec \text{dequeue}(Q_w) \\
\forall w \in W, \forall l' \in \text{dequeue}(Q_w) : \text{dequeue}(Q_w) \prec \text{execute-all-abilities}(w, l')
\end{array}
\]

where \( \prec \) denotes execution precedence, \( l \) represents either a node or edge location, and \newline
\(\text{execute-all-abilities}(w, l)\) represents the complete sequence of ability executions for walker \(w\) at location \(l\).

This execution model creates a bidirectional coupling between data and computation that is central to the OSP paradigm, allowing both locations and walkers to respond to traversal events in a coordinated sequence while maintaining the queue-based traversal mechanism that guides walkers through the topological structure.

\subsubsection{Self and Contextual References}
\label{subsubsec:selfhere}

To support the implicit execution model of abilities, OSP provides special reference mechanisms that give abilities access to their execution context:

\begin{itemize}
    \item \textbf{Self-reference} (\(\mathbf{self}\)): Traditional self-reference within an instance, providing access to the instance's own properties and methods. In walker abilities, \(\mathbf{self}\) refers to the walker instance, while in node or edge abilities, \(\mathbf{self}\) refers to the node or edge instance respectively.
    
    \item \textbf{Here-reference} (\(\mathbf{here}\)): In walker abilities, \(\mathbf{here}\) refers to the current location (node or edge) the walker is positioned at, providing access to the location's properties and methods from the walker's perspective. This enables walkers to interact with their current object-spatial context, representing mobile computation accessing local data.

    \item \textbf{Visitor-reference} (\(\mathbf{visitor}\)): In node or edge abilities, \(\mathbf{visitor}\) refers to the walker triggering the ability, providing access to the walker's properties and methods from the location's perspective. This enables locations to interact with visiting walkers based on their specific properties, representing data-bound computation accessing the mobile computational entity.

    \item \textbf{Path-reference} (\(\mathbf{path}\)): In all abilities, \(\mathbf{path}\) provides access to the walker's traversal path and destination queue \(Q_w\), allowing inspection and manipulation of the planned traversal sequence. This enables walkers to dynamically adjust their future movement based on conditions encountered during traversal. The path reference gives runtime access to the entire destination queue, enabling operations such as:
        \begin{itemize}
            \item Viewing the next planned destinations
            \item Modifying the order of destinations
            \item Inserting or removing destinations based on runtime conditions
            \item Querying path properties such as length, connectivity, or destination types
        \end{itemize}
\end{itemize}

These contextual references create a dual-perspective model where both walkers and locations can access each other's state when they interact. This bidirectional access pattern enables rich interaction models where both entities can influence each other during traversal events, actualizing the tight coupling between data and computation that characterizes the OSP paradigm.

\subsection{Complete Topological Structure}
\label{subsec:completetopology}

With the foundational elements and constructs defined, we can now formally describe the complete topological structure that embodies a Object-Spatial Program. This structure encapsulates both the interconnected elements and the distributed computational capabilities:

\[
G = (N, E, W, Q, L)
\]

where:
\begin{itemize}
    \item \( N = \{n_1, n_2, \ldots, n_m\} \) is the set of all node instances, each representing both a data location and a potential site of computation in the topological space
    \item \( E = \{e_1, e_2, \ldots, e_k\} \) is the set of all edge instances, each defining not only connectivity between nodes but also computational behaviors at transitions, where each \( e_i = (n_{\text{src}}, n_{\text{dst}}) \)
    \item \( W = \{w_1, w_2, \ldots, w_j\} \) is the set of all walker instances, representing autonomous computational entities that activate location-bound behaviors
    \item \( Q = \{Q_{w_1}, Q_{w_2}, \ldots, Q_{w_j}\} \) is the set of all walker destination queues, representing the planned traversal sequences for each active walker
    \item \( L: W \rightarrow N \cup E \cup \{\emptyset\} \) is a location mapping function that associates each walker with its current position in the topology, where \(\emptyset\) indicates an inactive walker
\end{itemize}

The state of the topological structure at any given moment during program execution is characterized by:

\begin{enumerate}
    \item \textbf{Distributed Computational Capacity}: Unlike traditional models where computation is centralized in functions, the OSP model distributes computational capabilities across all elements:
    \begin{itemize}
        \item Nodes contain both data and computational abilities (\( a_{\text{node}} \)) that activate in response to walker visits
        \item Edges contain both relational data and computational abilities (\( a_{\text{edge}} \)) that execute during transitions
        \item Walkers contain both traversal logic and computational abilities (\( a_{\text{walker}} \)) that respond to encountered locations
    \end{itemize}
    
    \item \textbf{Static Topological State}: The configuration of nodes and edges that form the underlying graph structure. This includes:
    \begin{itemize}
        \item The set of all nodes \( N \) with their internal data states and latent computational abilities
        \item The set of all edges \( E \) with their connection patterns and transition-specific abilities
        \item The resulting graph connectivity properties, such as reachability between nodes
    \end{itemize}
    
    \item \textbf{Dynamic Computational State}: The current positions and planned movements of walkers within the structure. This includes:
    \begin{itemize}
        \item The location mapping \( L \) that tracks the current position of each walker
        \item The traversal queues \( Q_{w_i} \) for each active walker \( w_i \), defining its future traversal path
    \end{itemize}
\end{enumerate}

As the program executes, this topological structure evolves through several key mechanisms:

\begin{enumerate}
    \item \textbf{Computational Activation}: The OSP model fundamentally inverts traditional computation:
    \begin{itemize}
        \item Rather than data moving to stationary computation (functions), walkers activate computation embedded within themselves and the data locations they visit
        \item Nodes and edges contain dormant computational abilities that are triggered by compatible walkers
        \item The interaction between a walker and its current location creates a dynamic computational context where both entities can affect each other
    \end{itemize}
    
    \item \textbf{Structural Modifications}: Changes to the node-edge graph through:
    \begin{itemize}
        \item Creation or deletion of nodes, affecting the set \( N \)
        \item Creation or deletion of edges, affecting the set \( E \) and the connectivity of the graph
    \end{itemize}
    
    \item \textbf{Path Construction and Utilization}: Although not part of the fundamental structure, path collections \( \mathcal{P} \) play an important role as programming constructs that:
    \begin{itemize}
        \item Define traversable routes through the topology
        \item Provide abstraction mechanisms for walker traversal patterns
        \item Serve as intermediaries between static structure and dynamic traversal
    \end{itemize}
    
    \item \textbf{Computational Movements}: Changes to the positions and traversal plans of walkers through:
    \begin{itemize}
        \item Activation of walkers via the spawn operator, transitioning walkers from inactive objects to active object-spatial entities positioned at specific locations
        \item Traversal between nodes and edges via the visit statement, updating the location mapping \( L \) and modifying walker destination queues \( Q_{w_i} \)
        \item At each traversal step, triggering a cascade of ability executions that constitute the actual computational work
        \item Termination of traversals through disengage statements, removing walkers from the active set
    \end{itemize}
    
    \item \textbf{State Transformations}: Changes to the internal states of elements through:
    \begin{itemize}
        \item Execution of node and edge abilities triggered by walker visits, allowing data locations to compute in response to visitors
        \item Execution of walker abilities triggered by encountered locations, allowing computational entities to respond to data contexts
        \item Bidirectional modification of properties through the \textbf{self} and \textbf{here} references
        \item Distributed effects as computation ripples through the topology via walker traversal
    \end{itemize}
\end{enumerate}

This complete topological structure \( G \) provides a unified mathematical representation of the OSP paradigm's distinctive approach: computation is not centralized in functions but distributed throughout a topological structure. Nodes and edges are not merely passive data containers but active computational sites that respond to walker visits. Walkers serve as both computational entities and activation mechanisms that trigger dormant abilities embedded within the topology.

The execution semantics defined in the following section operate within this distributed computational framework, specifying precisely how the interaction between walkers and locations triggers computation throughout the system, creating a model where the boundary between data and computation is fundamentally blurred.
\section{Object-Spatial Execution Semantics}

The execution model of Object-Spatial Programming combines traditional method invocation with object-spatial traversal operations and context-sensitive execution. This section details how instances are created and how computation flows through the topological structure, fundamentally inverting the traditional relationship where data is moved to computation.

\subsection{Instantiation Rules}
\label{subsubsec:instantiation}

To maintain object-spatial graph consistency and support higher-order topological structures, OSP enforces specific instantiation constraints for different archetypes and references:

\begin{enumerate}
    \item \textbf{Object Instantiation}: Standard objects follow traditional OOP instantiation patterns, with constructors defining initial state.
    
    \item \textbf{Node Instantiation}: Nodes are instantiated like standard objects but gain the additional capability to serve as endpoints for edges and hosts for walkers. Their constructors may initialize object-spatial properties and connection capabilities. Nodes effectively become locations where data resides and computation can be triggered, rather than passive data containers.
    
    \item \textbf{Edge Instantiation}: An instance \( e \) of an edge class \( \tau_{\text{edge}} \) can only be created if two nodes \( n_{\text{src}}, n_{\text{dst}} \) exist and are specified upon instantiation. This constraint ensures that edges always connect existing object-spatial elements, preventing dangling connections and maintaining referential integrity within the topological structure.
    
    \item \textbf{Walker Instantiation}: An instance \( w \) of a walker class \( \tau_{\text{walker}} \) can be instantiated as a standard object without an initial location. In this state, the walker functions as a regular object with all its properties and methods accessible, but it does not participate in object-spatial traversal until activated via the spawn operator. 
\end{enumerate}

These instantiation rules ensure that all object-spatial elements—from individual nodes and edges to higher-order path collections—maintain topological consistency while providing flexible construction mechanisms. The rules for path collections are particularly important as they bridge between the static topological structure and the dynamic execution patterns of walkers, allowing for complex traversal strategies to be expressed concisely while preserving the integrity of the object-spatial model.

\subsection{Lifecycle Management}
\label{subsubsec:lifecycle}

OSP extends traditional object lifecycle management with specialized rules for object-spatial archetypes:

\begin{enumerate}

\item \textit{Object Lifecycle}: Standard object instances follow traditional object lifecycle patterns from OOP, with standard creation, usage, and garbage collection.
    
\item \textit{Walker Lifecycle}: Walkers have a dual lifecycle, existing first as standard objects and then potentially transitioning to active object-spatial entities through the spawn operator. When active within the topological structure, walkers maintain their position and traversal state. They can be deactivated and return to standard object status under program control or when their traversal completes. This lifecycle reflects the mobile nature of computation in OSP, where algorithmic behaviors physically move through the data topology.
    
\item \textit{Node Lifecycle}: When a node instance is deleted, all edge instances that connect to or from that node are automatically deleted as well. This cascading deletion ensures object-spatial integrity by preventing dangling edges that would otherwise reference non-existent nodes. This constraint is expressed formally as:
    
    \[
    \forall e \in \tau_{\text{edge}} \text{ where } e = (n_{\text{src}}, n_{\text{dst}}) : \text{del}(n_{\text{src}}) \lor \text{del}(n_{\text{dst}}) \Rightarrow \text{del}(e)
    \]
    
\item \textit{Edge Lifecycle}: Edge instances exist as long as both their source and destination nodes exist. They are automatically garbage collected when either endpoint node is deleted, or when explicitly deleted by the program.
\end{enumerate}

This lifecycle management system ensures that the topological structure remains consistent throughout program execution, with automatic cleanup of dependent connections when nodes are removed, while providing flexibility for walker activation and deactivation.

\subsection{Spawn Operator (\(\Rightarrow\))}
\label{subsubsec:spawn}

The \textbf{spawn operator} (\(\Rightarrow\)) activates a walker within the topological structure by placing it at a specified node, edge, or path. This operation transitions the walker from a standard object state to an active object-spatial entity within the graph \(G\):

\paragraph{Spawning on a Single Element}

For spawning on a node:
\[
w \Rightarrow n \rightarrow w'
\]

where:
\begin{itemize}
    \item \( w \) is a walker instance currently in an inactive state (\(L(w) = \emptyset\))
    \item \( n \in N \) is the node where the walker will be spawned
    \item \( w' \) is the resulting active walker with updated location \(L(w') = n\)
    \item \( Q_{w'} = [] \) (empty queue) as no further destinations are automatically queued
\end{itemize}

For spawning on an edge:
\[
w \Rightarrow e \rightarrow w'
\]

where:
\begin{itemize}
    \item \( w \) is a walker instance currently in an inactive state (\(L(w) = \emptyset\))
    \item \( e = (n_{\text{src}}, n_{\text{dst}}) \in E \) is the edge where the walker will be spawned
    \item \( w' \) is the resulting active walker with updated location \(L(w') = e\)
    \item \( Q_{w'} = [n_{\text{dst}}] \) by default, as edges cannot be terminal locations and require a subsequent node destination
\end{itemize}

\paragraph{Spawning on a Path}

The spawn operator can also be applied to a path collection, allowing a walker to begin traversing a connected substructure:

\[
w \Rightarrow \mathcal{P} \rightarrow w'
\]

where:
\begin{itemize}
    \item \( w \) is a walker instance currently in an inactive state (\(L(w) = \emptyset\))
    \item \( \mathcal{P} = [p_1, p_2, \ldots, p_k] \) is a path collection where each \(p_i \in N \cup E\)
    \item \( w' \) is the resulting active walker with updated location \(L(w') = p_1\)
    \item The remaining elements are directly added to the walker's queue: \(Q_{w'} = [p_2, \ldots, p_k]\)
    \item Since \(\mathcal{P}\) is well-formed by definition, the path maintains topological validity, ensuring the walker can traverse from each element to the next without additional path resolution
\end{itemize}

This  approach creates a direct mapping between the path collection and the walker's traversal queue, reflecting the assumption that path collections are already topologically valid. The spawn operation initiates the walker's journey through the specified path, activating it at the first element and scheduling visits to all subsequent elements in the exact order provided.

The spawn operation has several important properties that affect the topological structure \(G = (N, E, W, Q, L)\):

\begin{enumerate}
    \item It can only be applied to a walker that is not already active within the topological structure (\(L(w) = \emptyset\))
    
    \item When executed, it modifies the location mapping function \(L\) to position the walker at the specified location (the first element of the spawn target)
    
    \item It initializes the walker's queue \(Q_w\) based on the spawn target:
    \begin{itemize}
        \item For a single node: \(Q_w = []\) (empty queue)
        \item For a single edge: \(Q_w = [n_{\text{dst}}]\) (destination node)
        \item For a path: \(Q_w = [p_2, \ldots, p_k]\) (remaining path elements)
    \end{itemize}
    
    \item The operation triggers all entry abilities associated with the walker's arrival at the spawn location:
    \begin{itemize}
        \item First, relevant location entry abilities for the arriving walker type
        \item Next, relevant walker entry abilities for the location type
    \end{itemize}
    
    \item After spawning, the walker maintains its active state until either:
    \begin{itemize}
        \item Its queue is exhausted and all abilities at its final location have executed, at which point it automatically returns to an inactive state
        \item It is explicitly terminated via a disengage statement
    \end{itemize}
    
    \item The spawn operator marks the transition of computation from a dormant state to an active participant in the distributed computational system
    
    \item When spawned on a path, the walker traverses the elements in the exact order specified, relying on the path's well-formed nature to maintain topological validity
\end{enumerate}

The path-based spawn operation extends the expressiveness of the OSP model by allowing walkers to be initialized with a complete traversal plan, rather than building the traversal dynamically through visit statements. This enables more declarative expression of algorithms that operate on connected substructures.

The spawn operator creates a clear separation between the initialization and activation phases of walker usage, allowing for complex setup before object-spatial traversal begins. It also marks the moment when the distributed computational model comes alive, with location-bound abilities and walker abilities beginning their interplay throughout the topological structure.

\subsection{Visit Statement (\(\triangleright\))}
\label{subsubsec:visit}

The \textbf{visit statement} (\(\triangleright\)) enables a walker to move between nodes and edges in the topological structure, representing the dynamic traversal capability that is central to the OSP model. This statement produces side effects by adding destinations to the walker's traversal queue and can be applied to traverse either a single element, multiple elements based on directional constraints, or an entire path collection.

\paragraph{Single Element Traversal}

For node to node traversal:
\[
w \triangleright n \rightarrow n
\]

where \( w = (n_{\text{curr}}) \) and \( n \in \tau_{\text{node}} \), meaning that the walker at node \( n_{\text{curr}} \) directly visits node \( n \). This requires that a direct edge connection exists between \( n_{\text{curr}} \) and \( n \), i.e., \( \exists e \in \tau_{\text{edge}} : e = (n_{\text{curr}}, n) \lor e = (n, n_{\text{curr}}) \).

For node to edge traversal:
\[
w \triangleright e \rightarrow \{e, n_{\text{next}}\}
\]

where \( e = (n_{\text{src}}, n_{\text{dst}}) \) and \( w = (n_{\text{curr}}) \), meaning that the walker at node \( n_{\text{curr}} \) moves to edge \( e \). This requires that \( n_{\text{curr}} \) is one of the endpoints of \( e \), i.e., \( n_{\text{curr}} = n_{\text{src}} \lor n_{\text{curr}} = n_{\text{dst}} \). Importantly, when a walker visits an edge, both the edge and its appropriate endpoint node are automatically queued in sequence:

\begin{itemize}
\item If \( n_{\text{curr}} = n_{\text{src}} \), then \( n_{\text{next}} = n_{\text{dst}} \)
\item If \( n_{\text{curr}} = n_{\text{dst}} \), then \( n_{\text{next}} = n_{\text{src}} \)
\end{itemize}

\paragraph{Edge Traversal Constraints}

When a walker is positioned on an edge, the visit statement cannot be called. Instead, after completing any edge abilities, the walker automatically transitions to the appropriate endpoint node that was queued during the preceding visit operation:

\[
w = (e) \text{ where } e = (n_{\text{src}}, n_{\text{dst}}) \Rightarrow n_{\text{next}}
\]

This constraint ensures that edges serve as transitions between nodes rather than locations where traversal decisions are made, maintaining the natural flow of the walker through the topological structure.

\paragraph{Path Traversal}

The visit statement can be applied to path collections, allowing a walker to traverse a connected substructure:

\[
w \triangleright \mathcal{P} \rightarrow \{p_1, p_2, ..., p_k\}
\]

where:
\begin{itemize}
    \item \( w \) is an active walker instance currently at node \( n_{\text{curr}} \)
    \item \( \mathcal{P} = [p_1, p_2, \ldots, p_k] \) is a path collection where each \(p_i \in N \cup E\)
    \item The path's first element \( p_1 \) must be directly reachable from the walker's current location:
    \begin{itemize}
        \item If \( p_1 \) is a node, either \( n_{\text{curr}} = p_1 \) or there exists an edge connecting \( n_{\text{curr}} \) and \( p_1 \)
        \item If \( p_1 \) is an edge, \( n_{\text{curr}} \) must be an endpoint of \( p_1 \)
    \end{itemize}
    \item All elements in the path are added to the walker's traversal queue in their path order: \(Q_w \leftarrow Q_w \cup [p_1, p_2, \ldots, p_k]\)
    \item Since \(\mathcal{P}\) is well-formed by definition, the path maintains topological validity, ensuring the walker can traverse from each element to the next without additional path resolution
\end{itemize}

Note that a visit statement can only be initiated from a node, not from an edge, in accordance with the edge traversal constraints.

The visit statement has several important properties:

\begin{enumerate}
    \item It is only valid if the walker is currently active and located at a node within the topological structure. Visit statements cannot be executed while a walker is on an edge.
    
    \item When executed from a node, the walker destination queues up all selected elements to visit after completing execution at its current location.
    
    \item When visiting an edge, both the edge and its appropriate endpoint node are automatically queued, ensuring that walkers always progress through the structure in a node-edge-node pattern.
    
    \item Once the walker completes executing all exit abilities at its current node, it will move to the first queued destination and trigger all relevant entry abilities at that new location.
    
    \item When on an edge, after executing all edge abilities, the walker automatically transitions to the queued endpoint node without requiring an explicit visit statement.
    
    \item This process continues recursively, with the walker moving through all queued locations until there are no more destinations in its queue.
    
    \item When the queue becomes empty after all abilities at the walker's current location have executed, the walker automatically transitions back to an inactive state.
    
    \item Multiple calls to the visit statement from a node append destinations to the walker's existing queue, allowing for dynamic construction of traversal paths during execution.
    
    \item When visiting a path, the walker enqueues all elements in the exact order specified in the path, relying on the path's well-formed nature to maintain topological validity.
    
    \item When a path collection is used with the visit statement, the first element of the path must be directly reachable from the walker's current position, enforcing a strict reachability constraint.
    
    \item When directly visiting a node from another node, the walker implicitly traverses the connecting edge, but does not trigger any edge abilities or perform any edge-specific processing. This provides a shorthand for node-to-node traversal when the intermediate edge context is not relevant to the algorithm.
    
    \item The model ensures that walkers never "get stuck" on edges, as they always automatically progress to nodes where further traversal decisions can be made.
\end{enumerate}

The path-based visit statement extends the expressiveness of the OSP model by allowing walkers to enqueue entire connected substructures for traversal in a single operation. This enables more concise expression of algorithms that operate on related sets of nodes or edges, such as graph search, path following, or subgraph processing.

By supporting node-to-node, node-to-edge, and path-based traversal patterns, the visit statement enables programmatic expression of complex traversal paths, allowing algorithms to navigate the topological structure in a controlled and semantically meaningful way. The automatic queuing of endpoint nodes when visiting edges ensures that walkers maintain the fundamental node-edge-node traversal pattern that reflects the object-spatial topology. The clear distinction that visit statements can only be executed from nodes, not edges, reinforces the concept that nodes are decision points in the traversal process, while edges are transitions between nodes. This embodiment of computation moving to data creates a fundamentally different programming model compared to conventional approaches where data is passed to stationary functions.

\subsection{Additional Flow Control Statements}
\label{subsubsec:flowcontrol}

To provide finer control over walker traversal execution, OSP includes two additional specialized flow control statements that operate within the context of object-spatial execution:

\paragraph{Skip Statement}

The \textbf{skip} statement allows a walker to immediately terminate execution at its current location and proceed to the next location in its traversal queue:

\[
\text{skip}(w) \Rightarrow L(w) \leftarrow \text{dequeue}(Q_w)
\]

where:
\begin{itemize}
    \item \( w \) is the active walker instance with current location \(L(w)\)
    \item \( Q_w \) is the walker's traversal queue with at least one queued location
\end{itemize}

When a skip statement is executed:
\begin{itemize}
    \item All remaining ability execution at the current location is immediately terminated
    \item Any exit abilities for the current location type are bypassed
    \item The walker immediately updates its position: \(L(w) \leftarrow \text{dequeue}(Q_w)\)
    \item Normal entry ability execution begins at the new location following the established order:
    \begin{itemize}
        \item First, relevant location entry abilities for the arriving walker type
        \item Next, relevant walker entry abilities for the location type
    \end{itemize}
\end{itemize}

The skip statement is analogous to the \textit{continue} statement in traditional loop constructs, allowing the walker to abort processing at the current location while continuing its overall traversal. This enables efficient implementation of conditional processing logic where certain nodes or edges might be examined but not fully processed based on their properties or the walker's state, providing fine-grained control over the distributed computation process.

\paragraph{Disengage Statement}

The \textbf{disengage} statement allows a walker to immediately terminate its entire object-spatial traversal and return to an inactive object state:

\[
\text{disengage}(w) \Rightarrow L(w) \leftarrow \emptyset, Q_w \leftarrow []
\]

where:
\begin{itemize}
    \item \( w \) is the active walker instance with current location \(L(w) \in N \cup E\)
\end{itemize}

When a disengage statement is executed:
\begin{itemize}
    \item All remaining ability execution at the current location is immediately terminated
    \item Any exit abilities for the current location type are bypassed
    \item The walker's traversal queue is cleared: \(Q_w \leftarrow []\)
    \item The walker's location is set to inactive: \(L(w) \leftarrow \emptyset\)
    \item The walker transitions from an active participant in the distributed computational system to an inactive object
    \item The walker retains all its properties and data accumulated during traversal
\end{itemize}

The disengage statement is analogous to the \textit{break} statement in traditional loop constructs, allowing the walker to completely exit the object-spatial execution context. This enables early termination of traversals when certain conditions are met, such as finding a target node, completing a computation, or encountering an error condition.

Together with the visit statement, these flow control statements provide essential mechanisms for implementing complex traversal algorithms where the path and processing logic may adapt dynamically based on discovered data or computed conditions within the topological structure. They offer precise control over both the walker's movement through the topology and its participation in the distributed computational process that characterizes the OSP model.
\section{Implementation Considerations}
A practical implementation of the Object-Spatial Programming model must address several important concerns that collectively guide the development of concrete OSP implementations in various programming languages. By addressing these aspects systematically, we can ensure that the theoretical OSP model translates into practical, efficient, and usable programming tools that effectively leverage the paradigm shift from moving data to computation to moving computation to data.
\subsection{\textbf{Type Safety}} The archetype system should be integrated with the host language's type system to ensure that object-spatial constraints (such as walker traversal rules) are checked at compile time when possible. This integration enables early detection of topology violations and provides developers with immediate feedback about invalid traversal patterns.
\subsection{\textbf{Concurrency}} In multi-walker scenarios, access to shared node and edge data must be properly synchronized to prevent race conditions. Implementations may adopt various concurrency models, from simple locking mechanisms to more sophisticated actor-based approaches. The "computation moves to data" paradigm creates new challenges and opportunities for parallel execution models, particularly when multiple walkers operate on overlapping sections of the topological structure.
\subsection{\textbf{Efficiency}} Naive implementations of walker traversal could lead to performance issues in large topological structures. Optimizations such as object-spatial indexing, traversal path caching, or parallel walker execution may be necessary for practical applications. The performance characteristics of mobile computation differ from traditional models and may require specialized optimization techniques, particularly for applications with complex traversal patterns or large data structures.
\subsection{\textbf{Integration}} OSP should be designed to complement rather than replace existing OOP mechanisms, allowing for gradual adoption and integration with legacy codebases. Hybrid approaches may be needed to bridge the paradigm gap between conventional data-to-computation models and OSP's computation-to-data approach. This enables incremental migration of existing systems toward object-spatial programming patterns while preserving investment in existing code.
\subsection{\textbf{Walker State Management}} The multi-state nature of walkers (as standard objects, active on nodes, or active on edges) requires careful state management to ensure consistency between these modes of operation, particularly when transitioning between states via spawn and disengage operations. Implementations must guarantee that walker state remains coherent throughout traversal operations and across transitions between active and inactive states.
\subsection{\textbf{Entry/Exit Ability Optimization}} Implementations should efficiently dispatch entry and exit abilities to minimize overhead during traversal operations, particularly in performance-critical applications. The implicit execution model of abilities requires careful design to maintain predictable performance characteristics. Techniques such as ability caching, ahead-of-time compilation, or just-in-time optimization may be necessary for high-performance systems.
\subsection{\textbf{Edge Transition Management}} The automatic queuing of edge destination nodes requires efficient handling of traversal queues and execution contexts to maintain predictable walker flow through the topological structure. This includes optimizing the mechanisms for determining the appropriate destination node based on traversal direction and managing the lifecycle of edge-bound computational processes.
\subsection{\textbf{Flow Control Semantics}} Implementations must handle visit, skip, and disengage statements efficiently, ensuring proper cleanup of execution contexts and maintaining the integrity of the walker's state during these control flow transitions. This is particularly important for complex traversal patterns where walkers may dynamically alter their paths based on discovered data or computational results.
\subsection{\textbf{Data Locality and Caching}} The computation-to-data paradigm can potentially improve data locality and cache efficiency compared to traditional approaches, but implementations must be designed to fully exploit these advantages, particularly when walkers traverse both nodes and edges. Memory layout strategies that co-locate related nodes and edges can significantly improve performance by reducing cache misses during traversal operations.
\section{Example -- A Object-Spatial Social Media Application}
\label{sec:jac-implementation}

This section examines a practical implementation of Object-Spatial Programming (OSP) concepts using Jac~\cite{jasecical,JaseciGitHub,JacLangWebsite}, a language that supersets Python and provides native support for the OSP paradigm. The example presents a Twitter-like social media application~\cite{Jaseci-Labs_littleX} that demonstrates how OSP's topological relationships and computation-to-data approach can be applied to real-world problems.

\subsection{Overview of the Jac Implementation}
\label{subsec:jac-overview}

The implementation leverages Jac's support for OSP archetypes to model a social media platform's core features. We observe the clear manifestation of the four key archetypes from OSP semantics:

\begin{enumerate}

    \item \textbf{Object Classes} (\(\tau_{\text{obj}}\)): Implemented as \lstinline[basicstyle=\fontsize{9}{10}]{obj} in Jac, exemplified by the \lstinline[basicstyle=\fontsize{9}{10}]{TweetInfo} class.
    \item \textbf{Node Classes} (\(\tau_{\text{node}}\)): Implemented as \lstinline[basicstyle=\fontsize{9}{10}]{node} in Jac, represented by \lstinline[basicstyle=\fontsize{9}{10}]{Profile}, \lstinline[basicstyle=\fontsize{9}{10}]{Tweet}, and \lstinline[basicstyle=\fontsize{9}{10}]{Comment}.
    \item \textbf{Edge Classes} (\(\tau_{\text{edge}}\)): Implemented as \lstinline[basicstyle=\fontsize{9}{10}]{edge} in Jac, representing relationships such as \lstinline[basicstyle=\fontsize{9}{10}]{Follow}, \lstinline[basicstyle=\fontsize{9}{10}]{Like}, and \lstinline[basicstyle=\fontsize{9}{10}]{Post}.
    \item \textbf{Walker Classes} (\(\tau_{\text{walker}}\)): Implemented as \lstinline[basicstyle=\fontsize{9}{10}]{walker} in Jac, with examples like \lstinline[basicstyle=\fontsize{9}{10}]{visit_profile} and \lstinline[basicstyle=\fontsize{9}{10}]{load_feed}.
\end{enumerate}

The implementation showcases OSP's fundamental inversion of the traditional data-to-computation paradigm, instead moving computation to data through walkers visiting nodes and triggering appropriate abilities.

\subsection{Core OSP Capabilities}
\label{subsec:core-capabilities}

\subsubsection{Object Definition}
\label{subsubsec:object-definition}

The implementation begins with a traditional object used for data transfer. The \lstinline[basicstyle=\fontsize{9}{10}]{TweetInfo} class exemplifies the standard object archetype (\(\tau_{\text{obj}}\)) as described in the OSP semantics. It encapsulates tweet metadata, content, and relationship data, serving as a non-topological data container that maintains compatibility with traditional object-oriented paradigms while enabling integration with OSP extensions.

This object is particularly significant as it bridges the OSP graph structure with traditional data representation needs, allowing data collected from the topological structure to be packaged in a format suitable for API responses and client-side processing.

\begin{lstlisting}[language=jac, caption={Object Definition in Jac}, label={lst:object-def}]
obj TweetInfo {
    has username: str;       # Author of the tweet
    has id: str;             # Unique identifier
    has content: str;        # Text content
    has embedding: list;     # Vector representation for search
    has likes: list;         # Users who liked the tweet
    has comments: list;      # Associated comments
}
\end{lstlisting}

\subsubsection{Node Definitions}
\label{subsubsec:node-definitions}

The application defines three node archetypes (\(\tau_{\text{node}}\)) representing the primary data entities in the social network. These nodes form the backbone of the social media platform, providing locations where data resides and computation occurs.

\paragraph{Profile Node}
The \lstinline[basicstyle=\fontsize{9}{10}]{Profile} node represents a user in the social network. It stores basic user information and implements abilities for account management and social connections. This node demonstrates how a object-spatial element can encapsulate both state (the username) and behavior (the follow/unfollow abilities).

Profile abilities showcase the implicit execution model of OSP. Rather than explicit method calls, computation is triggered automatically when appropriate walkers enter or exit the node. For example, the \lstinline[basicstyle=\fontsize{9}{10}]{update} ability is triggered when an \lstinline[basicstyle=\fontsize{9}{10}]{update_profile} walker enters the node, while the \lstinline[basicstyle=\fontsize{9}{10}]{follow} ability creates a social connection when a \lstinline[basicstyle=\fontsize{9}{10}]{follow_request} walker arrives.

\begin{lstlisting}[language=jac, caption={Profile Node Definition}, label={lst:profile-node}]
node Profile {
    has username: str = "";

    can update with update_profile entry {
        self.username = here.new_username;
        report self;
    }

    can get with get_profile entry {
        follwers=[{"id": jid(i), "username": i.username} for i in [self-->(`?Profile)]];
        report {"user": self, "followers": follwers};
    }

    can follow with follow_request entry {
        current_profile = [root-->(`?Profile)];
        current_profile[0] +:Follow():+> self;
        report self;
    }

    can un_follow with un_follow_request entry {
        current_profile = [root-->(`?Profile)];
        follow_edge = :e:[current_profile[0] -:Follow:-> self];
        Jac.destroy(follow_edge[0]);
        report self;
    }
}
\end{lstlisting}

\paragraph{Tweet Node}
The \lstinline[basicstyle=\fontsize{9}{10}]{Tweet} node represents a single post in the social network. It stores the content of the post, its semantic embedding for search, and a timestamp. This node demonstrates the richest set of abilities in the application, handling content updates, likes, comments, and query-based retrieval.

The Tweet node exemplifies how OSP nodes can act as both data containers and computational units. The \lstinline[basicstyle=\fontsize{9}{10}]{get_info} method is a traditional function that processes and returns data, while abilities like \lstinline[basicstyle=\fontsize{9}{10}]{like_tweet} and \lstinline[basicstyle=\fontsize{9}{10}]{comment} modify the topological structure by creating new edges and nodes.

Particularly significant is the \lstinline[basicstyle=\fontsize{9}{10}]{get} ability triggered by the \lstinline[basicstyle=\fontsize{9}{10}]{load_feed} walker, which calculates semantic similarity between a search query and the tweet content, demonstrating how OSP can combine graph traversal with computational operations.

\begin{lstlisting}[language=jac, caption={Tweet Node Definition}, label={lst:tweet-node}]
node Tweet {
    has content: str;
    has embedding: list;
    has created_at: str = datetime.datetime.now().strftime("%Y-%m-%d %H:%M:%S");

    can update with update_tweet exit {
        self.content = here.updated_content;
        report self;
    }

    can delete with remove_tweet exit {
        Jac.destroy(self);
    }

    can like_tweet with like_tweet entry {
        current_profile = [root-->(`?Profile)];
        self +:Like():+> current_profile[0];
        report self;
    }

    can remove_like with remove_like entry {
        current_profile = [root-->(`?Profile)];
        like_edge = :e:[self -:Like:-> current_profile[0]];
        Jac.destroy(like_edge[0]);
        report self;
    }

    can comment with comment_tweet entry {
        current_profile = [root-->(`?Profile)];
        comment_node = current_profile[0] ++> Comment(content=here.content);
        Jac.unrestrict(comment_node[0], level="CONNECT");
        self ++> comment_node[0];
        report comment_node[0];
    }

    can get_info()-> TweetInfo {
        return TweetInfo(
            username=[self<-:Post:-][0].username,
            id=jid(self),
            content=self.content,
            embedding=self.embedding,
            likes=[i.username for i in [self-:Like:->]],
            comments=[{"username": [i<--(`?Profile)][0].username, "id": jid(i), "content": i.content} 
                     for i in [self-->(`?Comment)]]
        );
    }

    can get with load_feed entry {
        tweet_info = self.get_info();
        similarity = search_tweets(here.search_query, tweet_info.content);
        here.results.append({"Tweet_Info": tweet_info, "similarity": similarity});
    }
}
\end{lstlisting}

\paragraph{Comment Node}
The \lstinline[basicstyle=\fontsize{9}{10}]{Comment} node represents user responses to tweets. As the simplest node type in the application, it primarily stores textual content and provides basic operations for updating and deleting. Comments form a secondary layer in the social graph, connected to both the originating tweet and the user who created them.

While minimal in structure, the Comment node completes the social interaction model by enabling multi-user discussions around content. Its connections in the graph (created by the Tweet node's \lstinline[basicstyle=\fontsize{9}{10}]{comment} ability) demonstrate how new topological elements can be dynamically added to extend the functionality of the system.

\begin{lstlisting}[language=jac, caption={Comment Node Definition}, label={lst:comment-node}]
node Comment {
    has content: str;

    can update with update_comment entry {
        self.content = here.updated_content;
        report self;
    }

    can delete with remove_comment entry {
        Jac.destroy(self);
    }
}
\end{lstlisting}

These node definitions collectively illustrate several key OSP concepts:

\begin{enumerate}
    \item \textbf{Data Encapsulation}: Each node type encapsulates specific data relevant to its role in the social network topology.
    
    \item \textbf{Ability Definition}: The nodes define abilities using the \lstinline[basicstyle=\fontsize{9}{10}]{can} keyword, which are automatically triggered based on walker interactions. These correspond directly to the OSP semantics of abilities (\(a_{\text{node}} : (\tau_{\text{walker}}, t) \rightarrow \bot\)) that execute without explicit parameters or return values.
    
    \item \textbf{Entry/Exit Abilities}: Each ability specifies whether it triggers on walker entry or exit using the \lstinline[basicstyle=\fontsize{9}{10}]{entry} and \lstinline[basicstyle=\fontsize{9}{10}]{exit} keywords, following the OSP ability execution model described in Section~\ref{subsubsec:abilityexec} of the OSP semantics.
    
    \item \textbf{Self/Here References}: The nodes use \lstinline[basicstyle=\fontsize{9}{10}]{self} to refer to their own state and \lstinline[basicstyle=\fontsize{9}{10}]{here} to refer to the visiting walker, precisely implementing the contextual reference mechanism described in Section~\ref{subsubsec:selfhere} of the OSP semantics.
\end{enumerate}

\subsubsection{Edge Definitions}
\label{subsubsec:edge-definitions}

The implementation defines three edge archetypes (\(\tau_{\text{edge}}\)) representing the core relationships in the social network. While these edge definitions appear minimal, they represent a powerful aspect of the OSP model: the elevation of relationships to first-class entities within the programming paradigm. 

Each edge type carries specific semantic meaning that shapes both the social graph structure and the allowed traversal patterns. The \lstinline[basicstyle=\fontsize{9}{10}]{Follow} edge creates the social connections between users that determine content visibility. The \lstinline[basicstyle=\fontsize{9}{10}]{Like} edge represents user interactions with content. The \lstinline[basicstyle=\fontsize{9}{10}]{Post} edge establishes authorship, connecting users to the content they create.

\begin{lstlisting}[language=jac, caption={Edge Definitions in Jac}, label={lst:edge-defs}]
edge Follow {}  # Follower -> followed relationship between profiles

edge Like {}    # Interaction between tweet and profile

edge Post {}    # Authorship connection from profile to tweet
\end{lstlisting}

Jac supports the edge semantics described in Section~\ref{subsubsec:formalization}, where an edge \(o_{\text{edge}} = (n_{\text{src}}, n_{\text{dst}})\) connects source and destination nodes. In Jac, the edge creation syntax explicitly specifies source and destination nodes, though in our updated OSP semantics, traversal direction is determined contextually during walker operations. The application demonstrates three distinct edge creation patterns:

\begin{itemize}
    \item \lstinline[basicstyle=\fontsize{9}{10}]{current_profile[0] +:Follow():+> self}: Creates a typed Follow edge with current\_profile[0] as the source node and self as the destination node, establishing a social connection.
    
    \item \lstinline[basicstyle=\fontsize{9}{10}]{self +:Like():+> current_profile[0]}: Creates a Like edge with the tweet (self) as the source node and current\_profile[0] as the destination node, recording a user interaction.
    
    \item \lstinline[basicstyle=\fontsize{9}{10}]{current_profile[0] ++> Comment(content=here.content)}: Creates both a new Comment node and an edge with current\_profile[0] as the source node connecting it to the newly created Comment node, demonstrating how topology can be dynamically extended.
\end{itemize}

The edges define the pathways along which walkers can traverse, serving not only as connections but also as potential locations where walkers can reside and perform computation. While Jac's syntax specifies source and destination nodes during edge creation, our updated OSP semantics would allow walkers to traverse these edges in either direction based on their current location, with the appropriate destination node determined by the traversal context. This implementation illustrates how even simple edge definitions enable complex social networking patterns when combined with appropriate node abilities and walker traversals.

\subsubsection{Walker Definitions}
\label{subsubsec:walker-definitions}

The implementation includes several walker archetypes (\(\tau_{\text{walker}}\)) that embody the computation-to-data paradigm central to OSP. Walkers represent the active computational elements that move through the topological structure, processing data and triggering behaviors as they visit different nodes and edges.

\paragraph{Profile Management Walkers}
The profile management walkers handle user account operations. The \lstinline[basicstyle=\fontsize{9}{10}]{visit_profile} walker serves as a base class that locates a user's profile or creates one if it doesn't exist. This demonstrates the OSP principle of traversal with conditional node creation. Other walkers like \lstinline[basicstyle=\fontsize{9}{10}]{update_profile} and \lstinline[basicstyle=\fontsize{9}{10}]{get_profile} extend this base walker, inheriting its traversal logic while adding specific data payloads.

The \lstinline[basicstyle=\fontsize{9}{10}]{load_user_profiles} walker demonstrates integration with traditional database queries, showing how OSP can complement conventional data access methods. Meanwhile, the minimal \lstinline[basicstyle=\fontsize{9}{10}]{follow_request} and \lstinline[basicstyle=\fontsize{9}{10}]{un_follow_request} walkers trigger specific node abilities when they arrive at a profile node.

\subsection{Advanced OSP Features}
\label{subsec:advanced-features}

\subsubsection{Topological Queries}
\label{subsubsec:topological-queries}

The implementation showcases Jac's support for topological queries, which allow for concise expression of graph traversal patterns. These queries align with the updated OSP semantics where traversal direction is determined contextually during the operation rather than being an intrinsic property of edges:

\begin{itemize}
    \item \lstinline[basicstyle=\fontsize{9}{10}]{[self-->(`?Profile)]}: Retrieves all Profile nodes connected by any outgoing edge from the current node's perspective. This directional constraint is applied at query time rather than being stored on the edge.
    
    \item \lstinline[basicstyle=\fontsize{9}{10}]{[current_profile[0] -:Follow:-> self]}: Finds edges of type Follow connecting the current profile to target profile, with direction specified as part of the traversal operation.
    
    \item \lstinline[basicstyle=\fontsize{9}{10}]{[-:Follow:->](`?Profile)}: Finds all Profile nodes connected by Follow edges that are outgoing from the current node's perspective.
\end{itemize}

These queries demonstrate how the topological relationships central to OSP can be efficiently queried with directional constraints applied during traversal. In Jac, the \lstinline[basicstyle=\fontsize{9}{10}]{-->} and \lstinline[basicstyle=\fontsize{9}{10}]{<--} operators specify traversal direction based on the edge's source and destination nodes as defined during creation. In our updated OSP semantics, these operators would map to directional selection based on the walker's current position relative to the edge endpoints, aligning with the \(w \triangleright \{e \in E | \text{dir}(e, n_{\text{curr}}) = d\}\) formalization.

\subsubsection{Dynamic Edge and Node Creation}
\label{subsubsec:dynamic-creation}

The implementation supports dynamic creation of both edges and nodes during walker traversal:

\begin{itemize}
    \item In the \lstinline[basicstyle=\fontsize{9}{10}]{follow} ability: \lstinline[basicstyle=\fontsize{9}{10}]{current_profile[0] +:Follow():+> self}
    \item In the \lstinline[basicstyle=\fontsize{9}{10}]{create_tweet} walker: \lstinline[basicstyle=\fontsize{9}{10}]{tweet_node = here +:Post:+> Tweet(content=self.content, embedding=embedding)}
    \item In the \lstinline[basicstyle=\fontsize{9}{10}]{comment} ability: \lstinline[basicstyle=\fontsize{9}{10}]{comment_node = current_profile[0] ++> Comment(content=here.content)}
\end{itemize}

This dynamic creation capability demonstrates the flexibility of the OSP model in evolving the topological structure during execution, rather than requiring a static graph structure defined at compile time. The syntax specifies source and destination nodes during edge creation, establishing a directed relationship between nodes. In our updated OSP semantics, while these relationships have defined source and destination nodes, walkers would be able to traverse edges in either direction based on their current location, with appropriate destination selection based on context.

\subsection{Case Study: Implementing the Feed Algorithm}
\label{subsec:case-study}

The implementation of a social media feed algorithm provides an exemplary demonstration of Object-Spatial Programming principles in action. This section analyzes how the \lstinline[basicstyle=\fontsize{9}{10}]{load_feed} walker and corresponding node abilities collaborate to implement a feature that demonstrates the OSP paradigm's distributed computation model.

\subsubsection{Algorithmic Components}
\label{subsubsec:algorithmic-components}

The feed algorithm comprises two complementary components that embody the OSP principle of distributing computation across the topological structure:

\begin{lstlisting}[language=jac, caption={Feed Algorithm Implementation in Walker}, label={lst:feed-algorithm}]
walker load_feed :visit_profile: {
    has search_query: str = "";  # Search term for filtering
    has results: list = [];      # Accumulates results

    can load with Profile entry {
        visit [-->(`?Tweet)];    # Visit user's own tweets
        
        for user_node in [-:Follow:->](`?Profile) {
            visit [user_node-->(`?Tweet)];  # Visit followed users' tweets
        }
        
        report self.results;     # Return collected results
    }
}
\end{lstlisting}

\begin{lstlisting}[language=jac, caption={Tweet Node Processing Logic}, label={lst:tweet-processing}]
can get with load_feed entry {
    tweet_info = self.get_info();
    similarity = search_tweets(here.search_query, tweet_info.content);
    here.results.append({"Tweet_Info": tweet_info, "similarity": similarity});
}
\end{lstlisting}

The implementation follows a two-phase traversal strategy that mirrors standard social media feed behavior:
\begin{enumerate}
    \item The walker first visits all tweets directly connected to the user's profile, using the directional traversal syntax \lstinline[basicstyle=\fontsize{9}{10}]{-->} to select edges where the profile is the source node
    \item It then traverses the social graph via \lstinline[basicstyle=\fontsize{9}{10}]{Follow} edges to access content from followed accounts, specifically following edges where the current profile is the source node and the followed profiles are destination nodes
\end{enumerate}

In Jac, these traversal operations use the \lstinline[basicstyle=\fontsize{9}{10}]{-->} and \lstinline[basicstyle=\fontsize{9}{10}]{<--} operators to follow edges based on their source and destination nodes as specified during creation. In our OSP semantics, these would be interpreted as directional selection operations similar to: \(w \triangleright \{e \in E | \text{dir}(e, n_{\text{curr}}) = d\}\), where outgoing edges would be those where the current node is the source, and incoming edges would be those where the current node is the destination.

\subsubsection{OSP-Centric Distribution of Responsibilities}
\label{subsubsec:responsibility-distribution}

The implementation demonstrates a clear separation of responsibilities that exemplifies the OSP paradigm's computation-to-data approach:

\paragraph{Traversal Logic (Walker)}
The \lstinline[basicstyle=\fontsize{9}{10}]{load_feed} walker encapsulates:
\begin{itemize}
    \item Profile resolution (inherited from its parent walker)
    \item Graph traversal strategy with context-sensitive directional constraints
    \item State management for query parameters and result accumulation
    \item Final result aggregation and reporting
\end{itemize}

\paragraph{Content Processing (Node)}
The Tweet node's \lstinline[basicstyle=\fontsize{9}{10}]{get} ability implements:
\begin{itemize}
    \item Self-contained tweet metadata collection
    \item Semantic similarity calculation against the search query
    \item Result contribution to the walker's accumulated state
\end{itemize}

This division implements the fundamental OSP principle identified in Section~\ref{sec:jac-implementation}: computation moves to data rather than data moving to computation. Each tweet processes itself in-place when visited, rather than being collected and processed by a centralized algorithm.

\subsubsection{Contextual Reference Mechanism}
\label{subsubsec:contextual-mechanism}

The implementation leverages OSP's contextual reference mechanism (as formalized in Section~\ref{subsubsec:selfhere}) to enable bidirectional information exchange between walker and node:

\begin{enumerate}
    \item \textbf{Walker-to-Node Data Flow}: The Tweet node accesses the search query parameter via \lstinline[basicstyle=\fontsize{9}{10}]{here.search_query}
    \item \textbf{Node-to-Walker Data Flow}: The Tweet node contributes to the result set via \lstinline[basicstyle=\fontsize{9}{10}]{here.results.append()}
\end{enumerate}

This mechanism creates a bridge between the traversal logic and processing logic without requiring explicit parameter passing or return values, implementing the ability execution semantics described in Section~\ref{subsec:abilities}. The same mechanism would apply when walkers visit edges, though this specific example does not demonstrate edge-specific abilities.

\subsubsection{OSP Principles Exemplified}
\label{subsubsec:dsp-principles}

This case study illustrates five fundamental OSP principles that characterize the paradigm:

\begin{enumerate}
    \item \textbf{Location-Based Computation}: Processing occurs at the data's residence rather than requiring data movement, reducing the need for intermediate data structures and minimizing state transfer.
    
    \item \textbf{Implicit Execution Model}: Node abilities activate automatically when corresponding walkers arrive, implementing the reactive execution model described in Section~\ref{subsubsec:abilityexec}.
    
    \item \textbf{Topological Algorithm Expression}: The social graph's structure directly influences the algorithm's behavior, with directed connections determining content visibility and accessibility based on the source-destination relationship of the edges.
    
    \item \textbf{State Transportation}: The walker functions as a mobile container that carries both input parameters and accumulates output as it traverses the topology.
    
    \item \textbf{Natural Distribution of Logic}: Responsibilities align naturally with their domain—traversal logic resides in walkers, while self-processing logic resides in nodes.
\end{enumerate}

\subsubsection{Architectural Advantages}
\label{subsubsec:architectural-advantages}

The OSP approach to implementing the feed algorithm offers several architectural benefits that address challenges commonly encountered in traditional paradigms:

\begin{itemize}
    \item \textbf{Encapsulation}: Tweet processing logic remains encapsulated within the Tweet node, providing a single source of truth for how tweets should process themselves.
    
    \item \textbf{Independent Evolution}: Traversal strategies can evolve independently from node processing logic, enabling separation of concerns across architectural boundaries.
    
    \item \textbf{Polymorphic Behavior}: Different node types can implement type-specific processing for the same walker type, enabling polymorphic behavior without complex conditional logic.
    
    \item \textbf{Scalability}: Processing is naturally distributed across the graph, potentially enabling parallel execution of node abilities in a distributed computing environment.
\end{itemize}

This implementation demonstrates how OSP provides a natural expression model for algorithms that operate over complex relational data structures. Rather than forcing either data-centric or algorithm-centric approaches, OSP enables a collaborative model where traversal agents interact with intelligent data nodes, creating systems that more faithfully represent domains with rich relationships and distributed processing requirements.
\section{Conclusion}
\label{sec:conclusion}

Object-Spatial Programming represents a principled application of the undecidability-lessening insight to the domain of topologically-structured computation. By elevating spatial and topological semantics to first-class language constructs through our four archetypes—object classes, node classes, edge classes, and walker classes—OSP fundamentally inverts the traditional relationship between data and computation. This paradigm shift from "data moving to computation" to "computation moving to data" transforms previously opaque program behaviors into explicit, observable patterns that enable sophisticated optimizations across the computing stack. Our formalization establishes rigorous semantics for instantiation, lifecycle management, and traversal operations while introducing abilities as spatially-triggered computational events that distribute processing throughout the topological structure. The resulting programming model provides natural abstractions for domains where connection topology is central—from social networks and agent-based systems to distributed computing and neural networks—while simultaneously enriching the semantic information available to compilers and runtime systems for automated optimization.

The significance of OSP extends beyond its immediate applications to demonstrate a broader design principle: that thoughtful abstraction design can simultaneously enhance programmer expressiveness and enable system-level optimizations that would be impossible with traditional programming models. As our social media case study illustrates, OSP enables more intuitive modeling of complex, interconnected systems while providing runtime environments with the semantic information necessary for intelligent data placement, parallel execution, and distribution strategies. Future research directions include developing sophisticated type systems for static verification of topological constraints, exploring advanced concurrency models for distributed walker execution, and investigating hybrid approaches that enable gradual adoption within existing codebases. As computational systems continue to grow in complexity and interconnectedness, programming paradigms that directly encode topological relationships and traversal patterns will become essential tools for managing this complexity while extracting maximum performance from modern hardware architectures.

\bibliographystyle{ACM-Reference-Format}
\bibliography{references}

\newpage
\appendix

\section{Structural Operational Semantics}

To formalize the execution model of Object-Spatial Programming, we present a set of structural operational semantics (SOS) rules that precisely define the state transitions that occur during program execution. These rules capture the core computational dynamics of OSP, with particular emphasis on spatial traversal and implicit ability execution.

\subsection{System State}

We first define the system state \( \sigma \) as a tuple:

\[
\sigma = (O, N, E, W, Q, L, A, S)
\]

where:
\begin{itemize}
    \item \( O \) is the set of all object instances in the system, where $O = \{o | o \text{ is an instance of } \tau_{\text{obj}}\}$.
    \item \( N \subset O \) is the set of all node instances, where $N = \{n | n \text{ is an instance of } \tau_{\text{node}}\}$.
    \item \( E \subset O \) is the set of all edge instances, where $E = \{e | e \text{ is an instance of } \tau_{\text{edge}}\}$ and each \( e \in E \) is of the form \( (n_{\text{src}}, n_{\text{dst}}) \) for $n_{\text{src}}, n_{\text{dst}} \in N$.
    \item \( W \subset O \) is the set of all walker instances, where $W = \{w | w \text{ is an instance of } \tau_{\text{walker}}\}$.
    \item \( Q : W \rightarrow \text{seq}(N \cup E) \) is a function mapping each walker to its destination queue, which is a sequence of nodes or edges to visit.
    \item \( L : W \rightarrow (N \cup E \cup \{\emptyset\}) \) is a function mapping each walker to its current location (node or edge), or \(\emptyset\) if the walker is inactive.
    \item \( A : W \rightarrow \{\text{active}, \text{inactive}\} \) is a function indicating whether each walker is active within the spatial structure.
    \item \( S : W \times E \rightarrow N \) is a source function mapping active walkers on edges to their entry node, tracking from which node the walker entered the edge.
\end{itemize}

This system state representation directly corresponds to the complete topological structure $G = (N, E, W, Q, L)$ defined in Section~\ref{subsec:completetopology}, with the addition of activity tracking $A$ and source tracking $S$ to facilitate operational semantics.

\subsection{Transition Rules}

We define the following transition rules that govern how the system state evolves during execution:

\subsubsection{Object Creation}
For a standard object construction:
\begin{align}
\frac{\text{new } \tau_{\text{obj}}()}{\sigma = (O, N, E, W, Q, L, A, S) \Rightarrow \sigma' = (O \cup \{o\}, N, E, W, Q, L, A, S)} \text{ [O-CREATION]}
\end{align}
where \( o \) is a fresh instance of type \( \tau_{\text{obj}} \).

\subsubsection{Node Creation}
For node construction:
\begin{align}
\frac{\text{new } \tau_{\text{node}}()}{\sigma = (O, N, E, W, Q, L, A, S) \Rightarrow \sigma' = (O \cup \{n\}, N \cup \{n\}, E, W, Q, L, A, S)} \text{ [N-CREATION]}
\end{align}
where \( n \) is a fresh instance of type \( \tau_{\text{node}} \).

\subsubsection{Edge Creation}
For edge construction between existing nodes, enforcing the instantiation constraints from Section~\ref{subsubsec:instantiation}:
\begin{align}
\frac{
\begin{array}{c}
\text{new } \tau_{\text{edge}}(n_{\text{src}}, n_{\text{dst}}) \\
n_{\text{src}}, n_{\text{dst}} \in N
\end{array}
}{\sigma = (O, N, E, W, Q, L, A, S) \Rightarrow \sigma' = (O \cup \{e\}, N, E \cup \{e\}, W, Q, L, A, S)} \text{ [E-CREATION]}
\end{align}
where \( e = (n_{\text{src}}, n_{\text{dst}}) \) is a fresh edge instance connecting nodes \( n_{\text{src}} \) and \( n_{\text{dst}} \).

\subsubsection{Walker Creation}
For walker construction in inactive state:
\begin{align}
&\frac{\text{new } \tau_{\text{walker}}()}{\sigma = (O, N, E, W, Q, L, A, S) \Rightarrow \sigma'} \text{ [W-CREATION]}\\
&\text{where } \sigma' = (O \cup \{w\}, N, E, W \cup \{w\}, Q[w \mapsto \langle\rangle], L[w \mapsto \emptyset], A[w \mapsto \text{inactive}], S) \nonumber
\end{align}
where \( w \) is a fresh walker instance, \( Q[w \mapsto \langle\rangle] \) assigns an empty queue to \( w \), \( L[w \mapsto \emptyset] \) indicates the walker has no location, and \( A[w \mapsto \text{inactive}] \) sets the walker's initial state to inactive, aligning with the inactive walker state defined in Section~\ref{subsubsec:formalization}.

\subsubsection{Node Deletion}
For node deletion with cascade to connected edges, implementing the lifecycle management from Section~\ref{subsubsec:lifecycle}:
\begin{align}
&\frac{
\begin{array}{c}
\text{delete}(n) \\
n \in N \\
E_n = \{e \in E \mid e = (n, n') \lor e = (n', n) \text{ for any } n' \in N\}
\end{array}
}{\sigma = (O, N, E, W, Q, L, A, S) \Rightarrow \sigma'} \text{ [N-DELETION]}\\
&\text{where } \sigma' = (O \setminus (\{n\} \cup E_n), N \setminus \{n\}, E \setminus E_n, W, Q', L', A', S') \nonumber
\end{align}
where:
\begin{itemize}
    \item \( E_n \) is the set of all edges connected to node \( n \)
    \item \( Q' \) is \( Q \) with all references to \( n \) and edges in \( E_n \) removed from all queues
    \item \( L' \) is \( L \) updated to reflect that walkers at \( n \) or any edge in \( E_n \) are now at \( \emptyset \) and inactive
    \item \( A' \) is \( A \) updated to mark those walkers as inactive
    \item \( S' \) is \( S \) with mappings for deleted edges removed
\end{itemize}

This rule implements the cascading deletion behavior described in Section \ref{subsubsec:lifecycle}, ensuring that all edges connected to a deleted node are automatically removed to maintain object-spatial integrity.

\subsubsection{Spawn Operation on Node}
For spawning a walker at a specific node, implementing the Spawn Operator semantics from Section \ref{subsubsec:spawn}:
\begin{align}
&\frac{
\begin{array}{c}
w \Rightarrow n \\
w \in W \\
n \in N \\
A(w) = \text{inactive}
\end{array}
}{\sigma = (O, N, E, W, Q, L, A, S) \Rightarrow \sigma'} \text{ [SPAWN-NODE]}\\
&\text{where } \sigma' = (O, N, E, W, Q[w \mapsto \langle\rangle], L[w \mapsto n], A[w \mapsto \text{active}], S) \nonumber
\end{align}
where \( L[w \mapsto n] \) updates walker \( w \)'s location to node \( n \), and \( A[w \mapsto \text{active}] \) activates the walker.

\subsubsection{Spawn Operation on Edge}
For spawning a walker on a specific edge with default direction (from source to destination):
\begin{align}
&\frac{
\begin{array}{c}
w \Rightarrow e \\
w \in W \\
e = (n_{\text{src}}, n_{\text{dst}}) \in E \\
A(w) = \text{inactive}
\end{array}
}{\sigma = (O, N, E, W, Q, L, A, S) \Rightarrow \sigma'} \text{ [SPAWN-EDGE]}\\
&\text{where } \sigma' = (O, N, E, W, Q[w \mapsto \langle n_{\text{dst}} \rangle], L[w \mapsto e], A[w \mapsto \text{active}], S[w, e \mapsto n_{\text{src}}]) \nonumber
\end{align}
where:
\begin{itemize}
    \item \( L[w \mapsto e] \) updates walker \( w \)'s location to edge \( e \)
    \item \( A[w \mapsto \text{active}] \) activates the walker
    \item \( S[w, e \mapsto n_{\text{src}}] \) records the entry node as the source node for the walker's edge traversal
    \item \( Q[w \mapsto \langle n_{\text{dst}} \rangle] \) automatically adds the destination node to the walker's queue
\end{itemize}

\subsubsection{Spawn Operation on Edge with Explicit Entry Node}
For spawning a walker on a specific edge with an explicitly specified entry node, supporting the directional edge traversal described in Section \ref{subsubsec:spawn}:
\begin{align}
&\frac{
\begin{array}{c}
w \Rightarrow (e, n_{\text{entry}}) \\
w \in W \\
e = (n_{\text{src}}, n_{\text{dst}}) \in E \\
n_{\text{entry}} \in \{n_{\text{src}}, n_{\text{dst}}\} \\
A(w) = \text{inactive}
\end{array}
}{\sigma = (O, N, E, W, Q, L, A, S) \Rightarrow \sigma'} \text{ [SPAWN-EDGE-DIR]}\\
&\text{where } \sigma' = (O, N, E, W, Q', L[w \mapsto e], A[w \mapsto \text{active}], S[w, e \mapsto n_{\text{entry}}]) \nonumber
\end{align}
where:
\begin{itemize}
    \item \( L[w \mapsto e] \) updates walker \( w \)'s location to edge \( e \)
    \item \( A[w \mapsto \text{active}] \) activates the walker
    \item \( S[w, e \mapsto n_{\text{entry}}] \) records the explicit entry node for the walker's edge traversal
    \item \( Q' = \begin{cases} 
        Q[w \mapsto \langle n_{\text{dst}} \rangle] & \text{if } n_{\text{entry}} = n_{\text{src}} \\
        Q[w \mapsto \langle n_{\text{src}} \rangle] & \text{if } n_{\text{entry}} = n_{\text{dst}}
    \end{cases} \) automatically adds the appropriate destination node to the walker's queue
\end{itemize}

\subsubsection{Spawn Operation on Path}
For spawning a walker on a path collection, as specified in Section \ref{subsubsec:pathcollection} and Section \ref{subsubsec:spawn}:
\begin{align}
&\frac{
\begin{array}{c}
w \Rightarrow \mathcal{P} \\
\mathcal{P} = [p_1, p_2, \ldots, p_k] \\
p_1 \in N \cup E \\
w \in W \\
A(w) = \text{inactive}
\end{array}
}{\sigma = (O, N, E, W, Q, L, A, S) \Rightarrow \sigma'} \text{ [SPAWN-PATH]}\\
&\text{where } \sigma' = \begin{cases}
\begin{aligned}
&(O, N, E, W, Q[w \mapsto \langle p_2, \ldots, p_k \rangle], \\
&\quad L[w \mapsto p_1], A[w \mapsto \text{active}], S)
\end{aligned} & \text{if } p_1 \in N \\[10pt]
\begin{aligned}
&(O, N, E, W, Q[w \mapsto \langle n_{\text{next}}, p_2, \ldots, p_k \rangle], \\
&\quad L[w \mapsto p_1], A[w \mapsto \text{active}], S[w, p_1 \mapsto n_{\text{src}}])
\end{aligned} & \text{if } p_1 \in E
\end{cases} \nonumber
\end{align}
where:
\begin{itemize}
    \item \( L[w \mapsto p_1] \) places the walker at the first element in the path
    \item \( A[w \mapsto \text{active}] \) activates the walker
    \item If \(p_1\) is a node, the remaining path elements are directly queued
    \item If \(p_1\) is an edge \(e = (n_{\text{src}}, n_{\text{dst}})\), the appropriate endpoint \(n_{\text{next}} = n_{\text{dst}}\) and the remaining path elements are queued
    \item \( S[w, p_1 \mapsto n_{\text{src}}] \) records the entry node if \(p_1\) is an edge
\end{itemize}

This unified path spawn operation accommodates both node and edge paths as described in Section \ref{subsubsec:pathcollection}.

\subsubsection{Visit Operation from Node to Node}
For a walker visiting a directly connected node, implementing the node-to-node traversal from Section \ref{subsubsec:visit}:
\begin{align}
&\frac{
\begin{array}{c}
w \triangleright n_{\text{next}} \\
L(w) = n_{\text{curr}} \in N \\
\exists e \in E : e = (n_{\text{curr}}, n_{\text{next}}) \lor e = (n_{\text{next}}, n_{\text{curr}}) \\
A(w) = \text{active}
\end{array}
}{\sigma = (O, N, E, W, Q, L, A, S) \Rightarrow \sigma'} \text{ [VISIT-NODE-TO-NODE]}\\
&\text{where } \sigma' = (O, N, E, W, Q[w \mapsto Q(w) \cdot \langle n_{\text{next}} \rangle], L, A, S) \nonumber
\end{align}
where \( Q[w \mapsto Q(w) \cdot \langle n_{\text{next}} \rangle] \) appends node \( n_{\text{next}} \) to walker \( w \)'s destination queue. This implements direct node-to-node traversal, which implicitly crosses the connecting edge without triggering edge abilities.

\subsubsection{Visit Operation from Node to Edge}
For a walker visiting an edge from a node, implementing the node-to-edge traversal from Section \ref{subsubsec:visit}:
\begin{align}
&\frac{
\begin{array}{c}
w \triangleright e \\
e = (n_{\text{src}}, n_{\text{dst}}) \in E \\
L(w) = n_{\text{curr}} \in N \\
n_{\text{curr}} \in \{n_{\text{src}}, n_{\text{dst}}\} \\
A(w) = \text{active} \\
n_{\text{dest}} = \begin{cases} 
        n_{\text{dst}} & \text{if } n_{\text{curr}} = n_{\text{src}} \\
        n_{\text{src}} & \text{if } n_{\text{curr}} = n_{\text{dst}}
    \end{cases}
\end{array}
}{\sigma = (O, N, E, W, Q, L, A, S) \Rightarrow \sigma'} \text{ [VISIT-NODE-TO-EDGE]}\\
&\text{where } \sigma' = (O, N, E, W, Q[w \mapsto Q(w) \cdot \langle e, n_{\text{dest}} \rangle], L, A, S) \nonumber
\end{align}
where \( Q[w \mapsto Q(w) \cdot \langle e, n_{\text{dest}} \rangle] \) appends both the edge \( e \) and its appropriate destination node to walker \( w \)'s queue. This implements the node-to-edge traversal semantics, where both the edge and its opposite endpoint are automatically queued.

\subsubsection{Visit Operation to Path}
For a walker visiting a path collection, implementing the path traversal semantics from Section \ref{subsubsec:visit}:
\begin{align}
&\frac{
\begin{array}{c}
w \triangleright \mathcal{P} \\
\mathcal{P} = [p_1, p_2, \ldots, p_k] \\
L(w) = n_{\text{curr}} \in N \\
A(w) = \text{active} \\
\text{Reachable}(n_{\text{curr}}, p_1)
\end{array}
}{\sigma = (O, N, E, W, Q, L, A, S) \Rightarrow \sigma'} \text{ [VISIT-PATH]}\\
&\text{where } \sigma' = (O, N, E, W, Q[w \mapsto Q(w) \cdot \langle p_1, p_2, \ldots, p_k \rangle], L, A, S) \nonumber
\end{align}
where:
\begin{itemize}
    \item \( \text{Reachable}(n_{\text{curr}}, p_1) \) means that either:
    \begin{itemize}
        \item If \(p_1\) is a node: \( n_{\text{curr}} = p_1 \) or there exists an edge connecting \( n_{\text{curr}} \) and \( p_1 \)
        \item If \(p_1\) is an edge: \( n_{\text{curr}} \) is an endpoint of \( p_1 \)
    \end{itemize}
    \item \( Q[w \mapsto Q(w) \cdot \langle p_1, p_2, \ldots, p_k \rangle] \) appends all elements in the path to the walker's queue
\end{itemize}

This unified path visit operation accommodates both node and edge elements within the path, supporting the general path collection definition from Section \ref{subsubsec:pathcollection}.

\subsubsection{Multiple Visit Operation by Direction}
For a walker visiting multiple edges based on directional filtering, supporting the directional traversal capabilities described in Section~\ref{subsubsec:visit}:
\begin{align}
&\frac{
\begin{array}{c}
w \triangleright \{e \in E | \text{dir}(e, n_{\text{curr}}) = d\} \\
L(w) = n_{\text{curr}} \in N \\
A(w) = \text{active} \\
d \in \{\text{outgoing}, \text{incoming}, \text{any}\} \\
E_d = \{e \in E | e \text{ connects to } n_{\text{curr}} \text{ and } \text{dir}(e, n_{\text{curr}}) = d\}
\end{array}
}{\sigma = (O, N, E, W, Q, L, A, S) \Rightarrow \sigma'} \text{ [VISIT-DIRECTION]}\\
&\text{where } \sigma' = (O, N, E, W, Q[w \mapsto Q(w) \cdot \langle E_d \rangle], L, A, S) \nonumber
\end{align}
where:
\begin{itemize}
    \item \( d \in \{\text{outgoing}, \text{incoming}, \text{any}\} \) specifies the direction filter
    \item \( \text{dir}(e, n_{\text{curr}}) \) evaluates to \(\text{outgoing}\) if \(n_{\text{curr}} = n_{\text{src}}\) for edge \(e = (n_{\text{src}}, n_{\text{dst}})\), and \(\text{incoming}\) if \(n_{\text{curr}} = n_{\text{dst}}\)
    \item \( Q[w \mapsto Q(w) \cdot \langle E_d \rangle] \) appends the sequence of filtered edges and their destination nodes to the walker's queue
\end{itemize}

\subsubsection{Walker Movement from Node to Edge}
For a walker moving from a node to an edge in its queue after completing all abilities at the current node:
\begin{align}
&\frac{
\begin{array}{c}
Q(w) = e \cdot Q' \\
L(w) = n_{\text{curr}} \in N \\
e = (n_{\text{src}}, n_{\text{dst}}) \in E \\
n_{\text{curr}} \in \{n_{\text{src}}, n_{\text{dst}}\} \\
A(w) = \text{active} \\
\text{AbilitiesComplete}(w, n_{\text{curr}})
\end{array}
}{\sigma = (O, N, E, W, Q, L, A, S) \Rightarrow \sigma'} \text{ [W-NODE-TO-EDGE]}\\
&\text{where } \sigma' = (O, N, E, W, Q[w \mapsto Q'], L[w \mapsto e], A, S[w, e \mapsto n_{\text{curr}}]) \nonumber
\end{align}
where:
\begin{itemize}
    \item \( Q[w \mapsto Q'] \) updates walker \( w \)'s queue by removing the first element
    \item \( L[w \mapsto e] \) updates \( w \)'s location to the edge \( e \)
    \item \( S[w, e \mapsto n_{\text{curr}}] \) records the node from which the walker entered the edge
    \item \(\text{AbilitiesComplete}(w, n_{\text{curr}})\) ensures all abilities at the current node have finished execution, following the ability execution order defined in Section~\ref{subsubsec:abilityexec}
\end{itemize}

\subsubsection{Walker Movement from Edge to Node}
For a walker moving from an edge to a node in its queue after completing all abilities at the current edge:
\begin{align}
&\frac{
\begin{array}{c}
Q(w) = n_{\text{next}} \cdot Q' \\
L(w) = e \in E \\
e = (n_{\text{src}}, n_{\text{dst}}) \\
n_{\text{next}} \in \{n_{\text{src}}, n_{\text{dst}}\} \\
n_{\text{next}} \neq S(w, e) \\
A(w) = \text{active} \\
\text{AbilitiesComplete}(w, e)
\end{array}
}{\sigma = (O, N, E, W, Q, L, A, S) \Rightarrow \sigma'} \text{ [W-EDGE-TO-NODE]}\\
&\text{where } \sigma' = (O, N, E, W, Q[w \mapsto Q'], L[w \mapsto n_{\text{next}}], A, S) \nonumber
\end{align}
where:
\begin{itemize}
    \item \( Q[w \mapsto Q'] \) updates walker \( w \)'s queue by removing the first element
    \item \( L[w \mapsto n_{\text{next}}] \) updates \( w \)'s location to the node \( n_{\text{next}} \)
    \item \(\text{AbilitiesComplete}(w, e)\) ensures all abilities at the current edge have finished execution
    \item \( n_{\text{next}} \neq S(w, e) \) ensures the walker is moving to the node opposite from where it entered the edge
\end{itemize}

\subsubsection{Edge Automatic Queue Update}
For a walker that entered an edge but doesn't have the destination node queued, implementing the automatic endpoint queueing described in Section~\ref{subsubsec:abilityexec}:
\begin{align}
&\frac{
\begin{array}{c}
L(w) = e \in E \\
e = (n_{\text{src}}, n_{\text{dst}}) \\
S(w, e) = n_{\text{entry}} \\
n_{\text{dest}} = \begin{cases} 
        n_{\text{dst}} & \text{if } n_{\text{entry}} = n_{\text{src}} \\
        n_{\text{src}} & \text{if } n_{\text{entry}} = n_{\text{dst}}
    \end{cases} \\
n_{\text{dest}} \notin Q(w) \\
A(w) = \text{active}
\end{array}
}{\sigma = (O, N, E, W, Q, L, A, S) \Rightarrow \sigma'} \text{ [EDGE-AUTO-QUEUE]}\\
&\text{where } \sigma' = (O, N, E, W, Q[w \mapsto Q(w) \cdot \langle n_{\text{dest}} \rangle], L, A, S) \nonumber
\end{align}
where the appropriate destination node is automatically appended to the walker's queue if not already present.

\subsubsection{Skip Operation}
For skipping the remaining execution at the current location and proceeding to the next location in the queue, implementing the skip statement from Section~\ref{subsubsec:flowcontrol}:
\begin{align}
&\frac{
\begin{array}{c}
\text{skip}(w) \\
Q(w) = l_{\text{next}} \cdot Q' \\
L(w) = l_{\text{curr}} \in N \cup E \\
A(w) = \text{active}
\end{array}
}{\sigma = (O, N, E, W, Q, L, A, S) \Rightarrow \sigma'} \text{ [SKIP]}\\
&\text{where } \sigma' = \begin{cases} 
(O, N, E, W, Q[w \mapsto Q'], L[w \mapsto l_{\text{next}}], A, S) & \text{if } l_{\text{next}} \in N \\
(O, N, E, W, Q[w \mapsto Q'], L[w \mapsto l_{\text{next}}], A, S[w, l_{\text{next}} \mapsto l_{\text{curr}}]) & \text{if } l_{\text{next}} \in E \text{ and } l_{\text{curr}} \in N \\
\text{Error} & \text{if } l_{\text{next}} \in E \text{ and } l_{\text{curr}} \in E
\end{cases} \nonumber
\end{align}
where the walker immediately moves to the next location in its queue without executing any remaining abilities, bypassing exit abilities at the current location.

\subsubsection{Disengage Operation}
For disengaging a walker from the spatial structure, implementing the disengage statement from Section~\ref{subsubsec:flowcontrol}:
\begin{align}
&\frac{
\begin{array}{c}
\text{disengage}(w) \\
L(w) \in N \cup E \\
A(w) = \text{active}
\end{array}
}{\sigma = (O, N, E, W, Q, L, A, S) \Rightarrow \sigma'} \text{ [DISENGAGE]}\\
&\text{where } \sigma' = (O, N, E, W, Q[w \mapsto \langle\rangle], L[w \mapsto \emptyset], A[w \mapsto \text{inactive}], S') \nonumber
\end{align}
where:
\begin{itemize}
    \item \( Q[w \mapsto \langle\rangle] \) clears the walker's queue
    \item \( L[w \mapsto \emptyset] \) removes the walker's location
    \item \( A[w \mapsto \text{inactive}] \) sets the walker to inactive
    \item \( S' \) is \( S \) with any mapping for \( w \) removed
\end{itemize}

\subsubsection{Queue Exhaustion}
For a walker that has completed execution at its current location and has an empty queue:
\begin{align}
&\frac{
\begin{array}{c}
Q(w) = \langle\rangle \\
L(w) = n \in N \\
A(w) = \text{active} \\
\text{AbilitiesComplete}(w, n)
\end{array}
}{\sigma = (O, N, E, W, Q, L, A, S) \Rightarrow \sigma'} \text{ [QUEUE-EXHAUSTION-NODE]}\\
&\text{where } \sigma' = (O, N, E, W, Q, L, A[w \mapsto \text{inactive}], S) \nonumber
\end{align}
where the walker remains at its current node but transitions to an inactive state as described in Section~\ref{subsubsec:abilityexec}.

\subsection{Ability Execution Rules}

We now define rules that govern the activation and execution of abilities, reflecting the execution order specified in Section~\ref{subsubsec:abilityexec}:

\subsubsection{Node Entry Ability Activation}
For node entry abilities activating when a walker arrives:
\begin{align}
\frac{
\begin{array}{c}
L(w) = n \in N \\
A(w) = \text{active} \\
w \in \tau_{\text{walker}} \\
a \in \text{entry\_abilities}(n) \\
\text{matches}(a, \tau_{\text{walker}}) \\
\text{NotExecuted}(a, n, w)
\end{array}
}{\sigma \Rightarrow \sigma \oplus \text{execute}(a, n, w)} \text{ [N-ENTRY-ABILITY]}
\end{align}
where \( \oplus \) represents state composition after ability execution, \( \text{matches}(a, \tau_{\text{walker}}) \) indicates that ability \( a \) matches the walker type, and \(\text{NotExecuted}(a, n, w)\) ensures the ability hasn't already executed for this walker-node pair. This implements the location entry ability execution described in Section~\ref{subsubsec:abilityexec}.

\subsubsection{Walker Entry Ability for Node Activation}
For walker entry abilities activating after node entry abilities have completed:
\begin{align}
\frac{
\begin{array}{c}
L(w) = n \in N \\
A(w) = \text{active} \\
n \in \tau_{\text{node}} \\
a \in \text{entry\_abilities}(w) \\
\text{matches}(a, \tau_{\text{node}}) \\
\text{NodeEntryComplete}(n, w) \\
\text{NotExecuted}(a, w, n)
\end{array}
}{\sigma \Rightarrow \sigma \oplus \text{execute}(a, w, n)} \text{ [W-ENTRY-NODE-ABILITY]}
\end{align}
where \(\text{NodeEntryComplete}(n, w)\) ensures all node entry abilities have executed before walker entry abilities begin. This implements the execution precedence from Section~\ref{subsubsec:abilityexec} where location entry abilities execute before walker entry abilities.

\subsubsection{Edge Entry Ability Activation}
For edge entry abilities activating when a walker arrives:
\begin{align}
\frac{
\begin{array}{c}
L(w) = e \in E \\
A(w) = \text{active} \\
w \in \tau_{\text{walker}} \\
a \in \text{entry\_abilities}(e) \\
\text{matches}(a, \tau_{\text{walker}}) \\
\text{NotExecuted}(a, e, w)
\end{array}
}{\sigma \Rightarrow \sigma \oplus \text{execute}(a, e, w)} \text{ [E-ENTRY-ABILITY]}
\end{align}
where the edge's entry abilities are triggered when a walker enters the edge. This follows the ability execution order outlined in Section~\ref{subsubsec:abilityexec}.

\subsubsection{Walker Entry Ability for Edge Activation}
For walker entry abilities activating after edge entry abilities have completed:
\begin{align}
\frac{
\begin{array}{c}
L(w) = e \in E \\
A(w) = \text{active} \\
e \in \tau_{\text{edge}} \\
a \in \text{entry\_abilities}(w) \\
\text{matches}(a, \tau_{\text{edge}}) \\
\text{EdgeEntryComplete}(e, w) \\
\text{NotExecuted}(a, w, e)
\end{array}
}{\sigma \Rightarrow \sigma \oplus \text{execute}(a, w, e)} \text{ [W-ENTRY-EDGE-ABILITY]}
\end{align}
where \(\text{EdgeEntryComplete}(e, w)\) ensures all edge entry abilities have executed before walker entry abilities begin.

\subsubsection{Walker Exit Ability from Node Activation}
For walker exit abilities activating when preparing to leave a node:
\begin{align}
\frac{
\begin{array}{c}
L(w) = n \in N \\
A(w) = \text{active} \\
Q(w) \neq \langle\rangle \\
n \in \tau_{\text{node}} \\
a \in \text{exit\_abilities}(w) \\
\text{matches}(a, \tau_{\text{node}}) \\
\text{EntryComplete}(n, w) \\
\text{NotExecuted}(a, w, n)
\end{array}
}{\sigma \Rightarrow \sigma \oplus \text{execute}(a, w, n)} \text{ [W-EXIT-NODE-ABILITY]}
\end{align}
where \(\text{EntryComplete}(n, w)\) ensures all entry abilities have executed before exit abilities begin. This implements the behavior where all entry abilities execute before any exit abilities.

\subsubsection{Node Exit Ability Activation}
For node exit abilities activating after walker exit abilities have completed:
\begin{align}
\frac{
\begin{array}{c}
L(w) = n \in N \\
A(w) = \text{active} \\
Q(w) \neq \langle\rangle \\
w \in \tau_{\text{walker}} \\
a \in \text{exit\_abilities}(n) \\
\text{matches}(a, \tau_{\text{walker}}) \\
\text{WalkerExitComplete}(w, n) \\
\text{NotExecuted}(a, n, w)
\end{array}
}{\sigma \Rightarrow \sigma \oplus \text{execute}(a, n, w)} \text{ [N-EXIT-ABILITY]}
\end{align}
where \(\text{WalkerExitComplete}(w, n)\) ensures all walker exit abilities have executed before node exit abilities begin. This implements the execution precedence where walker exit abilities execute before location exit abilities.

\subsubsection{Walker Exit Ability from Edge Activation}
For walker exit abilities activating when preparing to leave an edge:
\begin{align}
\frac{
\begin{array}{c}
L(w) = e \in E \\
A(w) = \text{active} \\
Q(w) \neq \langle\rangle \\
e \in \tau_{\text{edge}} \\
a \in \text{exit\_abilities}(w) \\
\text{matches}(a, \tau_{\text{edge}}) \\
\text{EntryComplete}(e, w) \\
\text{NotExecuted}(a, w, e)
\end{array}
}{\sigma \Rightarrow \sigma \oplus \text{execute}(a, w, e)} \text{ [W-EXIT-EDGE-ABILITY]}
\end{align}
where \(\text{EntryComplete}(e, w)\) ensures all entry abilities have executed before exit abilities begin.

\subsubsection{Edge Exit Ability Activation}
For edge exit abilities activating after walker exit abilities have completed:
\begin{align}
\frac{
\begin{array}{c}
L(w) = e \in E \\
A(w) = \text{active} \\
Q(w) \neq \langle\rangle \\
w \in \tau_{\text{walker}} \\
a \in \text{exit\_abilities}(e) \\
\text{matches}(a, \tau_{\text{walker}}) \\
\text{WalkerExitComplete}(w, e) \\
\text{NotExecuted}(a, e, w)
\end{array}
}{\sigma \Rightarrow \sigma \oplus \text{execute}(a, e, w)} \text{ [E-EXIT-ABILITY]}
\end{align}
where \(\text{WalkerExitComplete}(w, e)\) ensures all walker exit abilities have executed before edge exit abilities begin.

\subsection{Contextual References}

The semantics of the special references \(\mathbf{self}\), \(\mathbf{here}\), \(\mathbf{visitor}\), and \(\mathbf{path}\) within ability execution contexts are defined as follows, implementing the contextual reference semantics from Section~\ref{subsubsec:selfhere}:

\begin{itemize}
    \item In a node ability \(a_{\text{node}}\) executing on node \(n\) triggered by walker \(w\):
    \begin{itemize}
        \item \(\mathbf{self} = n\) (the node instance itself)
        \item \(\mathbf{visitor} = w\) (the visiting walker)
        \item \(\mathbf{path} = Q_w\) (the walker's destination queue)
    \end{itemize}
    
    \item In an edge ability \(a_{\text{edge}}\) executing on edge \(e\) triggered by walker \(w\):
    \begin{itemize}
        \item \(\mathbf{self} = e\) (the edge instance itself)
        \item \(\mathbf{visitor} = w\) (the traversing walker)
        \item \(\mathbf{path} = Q_w\) (the walker's destination queue)
    \end{itemize}
    
    \item In a walker ability \(a_{\text{walker}}\) executing on walker \(w\) triggered at location \(l\) (where \(l\) is either a node or edge):
    \begin{itemize}
        \item \(\mathbf{self} = w\) (the walker instance itself)
        \item \(\mathbf{here} = l\) (the current node or edge location)
        \item \(\mathbf{path} = Q_w\) (the walker's own destination queue)
    \end{itemize}
\end{itemize}

These contextual references create the multi-perspective model described in Section \ref{subsubsec:selfhere}, where walkers, nodes, and edges can access each other's state during interaction, and all entities can inspect and potentially modify the walker's traversal path. This enables the rich bidirectional access pattern central to the OSP model.

\subsection{Traversal Direction Resolution}

The directional semantics of edge traversal, as used in rules like [VISIT-DIRECTION], are defined formally as follows:

\begin{enumerate}
    \item \textbf{Direction Evaluation Function}: The function $\text{dir}(e, n)$ evaluates the direction of edge $e = (n_{\text{src}}, n_{\text{dst}})$ relative to node $n$:
    \[
    \text{dir}(e, n) = 
    \begin{cases}
    \text{outgoing} & \text{if } n = n_{\text{src}} \\
    \text{incoming} & \text{if } n = n_{\text{dst}} \\
    \text{undefined} & \text{otherwise}
    \end{cases}
    \]
    
    \item \textbf{Traversal Path Resolution}: When a walker visits an edge from a node, the next node in its path is determined by edge orientation relative to the current node:
    \[
    \text{nextNode}(e, n_{\text{curr}}) = 
    \begin{cases}
    n_{\text{dst}} & \text{if } n_{\text{curr}} = n_{\text{src}} \\
    n_{\text{src}} & \text{if } n_{\text{curr}} = n_{\text{dst}}
    \end{cases}
    \]
    
    \item \textbf{Directional Selection}: The visit statement with directional filtering enables walkers to select edges based on their orientation:
    \[
    E_{\text{filtered}} = \{e \in E \mid e \text{ connects to } n_{\text{curr}} \text{ and } \text{dir}(e, n_{\text{curr}}) \in F\}
    \]
    where $F \subseteq \{\text{outgoing}, \text{incoming}, \text{any}\}$ is the set of desired directions.
\end{enumerate}

This contextual direction resolution enables the flexible traversal patterns described in Section~\ref{subsubsec:visit}, allowing walkers to navigate the topological structure in semantically meaningful ways.

\subsection{Execution Order Properties}

The operational semantics enforce the following essential ordering properties for ability execution, directly implementing the execution order from Section~\ref{subsubsec:abilityexec}:

\begin{enumerate}
    \item When a walker arrives at a node, node entry abilities execute before walker entry abilities:
    \[
    \forall a_{\text{node}}^{\text{entry}} \in n, \forall a_{\text{walker}}^{\text{entry}} \in w : \text{execute}(a_{\text{node}}^{\text{entry}}) \prec \text{execute}(a_{\text{walker}}^{\text{entry}})
    \]
    
    \item When a walker arrives at an edge, edge entry abilities execute before walker entry abilities:
    \[
    \forall a_{\text{edge}}^{\text{entry}} \in e, \forall a_{\text{walker}}^{\text{entry}} \in w : \text{execute}(a_{\text{edge}}^{\text{entry}}) \prec \text{execute}(a_{\text{walker}}^{\text{entry}})
    \]
    
    \item When a walker prepares to leave a node, walker exit abilities execute before node exit abilities:
    \[
    \forall a_{\text{walker}}^{\text{exit}} \in w, \forall a_{\text{node}}^{\text{exit}} \in n : \text{execute}(a_{\text{walker}}^{\text{exit}}) \prec \text{execute}(a_{\text{node}}^{\text{exit}})
    \]
    
    \item When a walker prepares to leave an edge, walker exit abilities execute before edge exit abilities:
    \[
    \forall a_{\text{walker}}^{\text{exit}} \in w, \forall a_{\text{edge}}^{\text{exit}} \in e : \text{execute}(a_{\text{walker}}^{\text{exit}}) \prec \text{execute}(a_{\text{edge}}^{\text{exit}})
    \]
    
    \item All entry abilities (both location and walker) execute before any exit abilities:
    \[
    \forall a^{\text{entry}}, \forall a^{\text{exit}} : \text{execute}(a^{\text{entry}}) \prec \text{execute}(a^{\text{exit}})
    \]
    
    \item The relationship between queue operations and ability execution follows this pattern:
    \[
    \begin{array}{c}
    \forall w \in W, \forall l \in L(w) : \text{execute-all-abilities}(w, l) \prec \text{dequeue}(Q_w) \\
    \forall w \in W, \forall l' \in \text{dequeue}(Q_w) : \text{dequeue}(Q_w) \prec \text{execute-all-abilities}(w, l')
    \end{array}
    \]
    where \( \prec \) denotes execution precedence, and \(\text{execute-all-abilities}(w, l)\) represents the complete sequence of ability executions for walker \(w\) at location \(l\).
\end{enumerate}

\subsection{Complete Execution Model}

The complete execution of a Object-Spatial Programming system follows these key principles, integrating all aspects of the OSP model from Sections~\ref{sec:semantics}-\ref{subsubsec:flowcontrol}:

\begin{enumerate}
    \item \textbf{System Initialization}: The execution begins with the initial state \( \sigma_0 \) representing an empty or pre-populated topological structure.
    
    \item \textbf{State Evolution}: At each step, applicable transition rules are applied to evolve the state \( \sigma_i \rightarrow \sigma_{i+1} \), representing the discrete computational steps of the OSP model.
    
    \item \textbf{Walker Activation and Movement}: Walkers transition between active and inactive states through spawn and disengage operations, and move through the topological structure via their destination queues, embodying the "computation moving to data" paradigm.
    
    \item \textbf{Strict Ability Execution Order}: When a walker arrives at a node, abilities execute in this strict order:
    \begin{enumerate}
        \item All node entry abilities matching the walker type
        \item All walker entry abilities matching the node type
        \item All walker exit abilities matching the node type (when preparing to leave)
        \item All node exit abilities matching the walker type (when preparing to leave)
    \end{enumerate}
    
    \item \textbf{Edge Transit Processing}: When a walker moves to an edge, abilities execute in this strict order:
    \begin{enumerate}
        \item All edge entry abilities matching the walker type
        \item All walker entry abilities matching the edge type
        \item The appropriate destination node is automatically added to the walker's queue if not already present
        \item All walker exit abilities matching the edge type (when preparing to leave)
        \item All edge exit abilities matching the walker type (when preparing to leave)
    \end{enumerate}
    
    \item \textbf{Coordinated Traversal Completion}: Walker movement occurs only after all triggered abilities have completed execution at the current location, unless explicitly skipped, ensuring coordinated activation of data-bound computation.
    
    \item \textbf{Path-Based Traversal}: Path collections provide a higher-order abstraction for expressing traversal patterns, with walkers automatically translating these paths into sequential traversal steps through their destination queues.
    
    \item \textbf{Bidirectional Computational Coupling}: The dual-perspective contextual reference system (self/here) creates a bidirectional coupling between walkers and locations, allowing both computational agents and data locations to respond to each other.
    
    \item \textbf{Termination Conditions}: The system continues to evolve until:
    \begin{itemize}
        \item All walkers have become inactive (either through queue exhaustion or explicit disengagement)
        \item A stable state is reached where no further transitions apply
        \item An infinite traversal pattern occurs (which would require external intervention or predefined termination criteria)
    \end{itemize}
\end{enumerate}

These operational semantics provide a formal foundation for reasoning about OSP programs, enabling rigorous analysis of:

\begin{itemize}
    \item \textbf{Program Correctness}: Verifying that traversal patterns and ability executions achieve desired computational outcomes
    \item \textbf{Termination Properties}: Ensuring that walker traversals eventually complete under expected conditions
    \item \textbf{Concurrency Behavior}: Analyzing potential race conditions or coordination issues when multiple walkers traverse the topology simultaneously
    \item \textbf{Execution Efficiency}: Optimizing traversal patterns and ability definitions for performance
    \item \textbf{Formal Verification}: Enabling formal proof techniques for safety and liveness properties
\end{itemize}

This formalization completes the mathematical foundation of Object-Spatial Programming, bridging the topological semantics of Section~\ref{sec:semantics} with the execution semantics of Section~\ref{subsubsec:flowcontrol} into a coherent computational model where the boundary between data and computation is fundamentally transformed.

\section{Complete Listing of Object-Spatial Twitter}


\pyinputlisting[language=jac, caption={Jac code from littlex.jac}, label={lst:littlex}]{littlex.jac}

\end{document}